\documentclass[prc, etal, amsmath,superscriptaddress,amssymb]{revtex4}

\usepackage{graphicx}
\usepackage{subfigure}
\usepackage{bm}
\usepackage{pgfplots}
\usepackage{mathtools}

\newcommand{\be}{\begin{equation}}
\newcommand{\ee}{\end{equation}}
\newcommand{\bea}{\begin{eqnarray}}
\newcommand{\eea}{\end{eqnarray}}

\newcommand{\gapp}{\mathrel{\raise.3ex\hbox{$>$}\mkern-14mu
\lower0.6ex\hbox{$\sim$}}}
\newcommand{\lapp}{\mathrel{\raise.3ex\hbox{$<$}\mkern-14mu
\lower0.6ex\hbox{$\sim$}}}
\def\bbox{{\,\lower0.9pt\vbox{\hrule \hbox{\vrule height 0.2 cm
\hskip 0.2 cm \vrule  height 0.2 cm}\hrule}\,}}

\begin{document}


\title{
Searching for Dark Matter Particles  with Compton Scattering Approach
}

\def\fdu {Institute of Modern Physics, Fudan University, Shanghai 200433, P.R. China}
\def\sjtu {School of Physics and Astronomy, Shanghai Jiao Tong University, Shanghai, 200240, China}
\def\iop {Laboratory of Optical Physics, Institute of Physics, Chinese Academy of Sciences, Beijing 100190, China}
\def\imp {Institute of Modern Physics, Chinese Academy of Sciences, Lanzhou, 730000, China}
\def\ciae {Department of Nuclear Physics, China Institute of Atomic Energy, Beijing, 102413, China}
\def\korea {Nuclear Data Center, Korea Atomic Energy Research Institute, Daejon 305353, Korea}
\def\nao {National Astronomical Observatories, Chinese Academy of Sciences, Beijing 100012, China}
\def\ifsa {IFSA Collaborative Innovation Center, Shanghai Jiao Tong University, Shanghai 200240, China}
\def\siom{Shanghai Institute of Optics and Fine Mechanics, Chinese Academy of Sciences, Shanghai 201800, China}
\def\SARI{Shanghai  Advanced Research Institute, Chinese Academy of Sciences, Shanghai 201210, China}
\def\LOP{Laboratory of Optical Physics, Institute of Physics, Chinese Academy of Sciences, Beijing 100190, China}
\def\UCA{School of Physical Sciences, University of Chinese Academy of Sciences, 100049 Beijing, China}
\def\YZU{ Center for Gravity and Cosmology, School of Phys. Sci. \&  Tech.,
Yangzhou University, 225002, China}
\def\case{ CERCA/Department of Physics/ISO, Case Western Reserve University, Cleveland OH 44106-7079}

\author{Shang Wang}\affiliation{\YZU}
\author{Changbo Fu}	\email[Corresponding author:] {cbfu@fudan.edu.cn} \affiliation{\fdu}
\author{De-Chang Dai}	\email[Corresponding author:]{diedachung@gmail.com}  \affiliation{\YZU}\affiliation{\case}
\author{Hongwei Wang} \affiliation{\SARI}
\author{Gongtao Fan} \affiliation{\SARI}
\author{Xiguang Cao} \affiliation{\SARI}
\author{Yugang Ma} \affiliation{\fdu}


\date{\today}

\begin{abstract}
The dark matter puzzle is one of the most important fundamental physics questions in 21 century.
There is no doubt that solving the puzzle will be a new milestone for human beings in the way of deeper understanding the mother nature.
Here we propose to use the Shanghai laser electron gamma source (SLEGS) to search for  dark matter candidates particles, including dark pseudo scalar particles, dark scalar particles, and dark photons.
Our simulations show that electron facilities like SLEGS with some upgrading could be competitive platforms in searching for light dark matter particles with mass under tens of keV.

\end{abstract}
\maketitle

\maketitle

\section{Introduction}
%
%
Whether dark matter particles (DMPs) exist or not is one of the biggest questions in modern physics\cite{DMP-RMP2018Bertone}.
Various cosmological and astrophysical observations show that invisible dark matter is responsible for more than 20\% of total energy in our universe\cite{WMAP2011}.
These experimental observations  include power spectra of cosmic microwave background\cite{WMAP2011,CMB1998}, 
rotation curves of galaxies\cite{glx-rot1996}, and gravitational lensing effects\cite{g-lensing2001} etc. 
Different experiments have been dedicated to searching for DMPs in broad ranges of mass, from sub-eV to TeV.
Candidates of DMPs include weakly interacting massive particles (WIMP), axion-like particle (ALP), dark photon, etc. \cite{DMP-Phys.Rep.BERTONE2005279}

Some DMPs, including scalar, pseudo scalar, or vector particles,  
may be generated in Compton processes where an electron collides with a photon\cite{DMP-Chakrabarty:2019kdd}.
For example, ALP, which is a pseudo scalar particle, is a promising  DMP candidate. 
Axion was first introduced by Wilczek and Weinberg in the 1960s as a result of spontaneously broken of the so-called Peccei-Quinn symmetry
\cite{axion-wilczek.1978, axion-Weinberg.1978, axion-PhysRevLett.38.1440}.
The axion provides a natural solution to the strong CP problem in QCD, as well as a good candidate for DMPs.
Many experimental methods, including helioscopes, light shining through a wall, microwave cavities, nuclear magnetic resonance, and the so-called axioelectrical effect,
have been taken to search for pseudo scalar particles (including axion or ALPs)\cite{raffelt2012limits,bulatowicz2013laboratory, ji2017searching,Ji.PRL2018}. 

Another DMP candidate is the dark photon\cite{DP-FILIPPI2020100042}. 
It arises from the symmetry of a hypothetical dark sector comprising particles completely neutral under the Standard Model interactions, which results in its darkness. Although its kinetic mixing with ordinary photons is very weak,  this new gauge boson may still be detectable. 
Scalar DMPs are also searched by many groups dedicatedly\cite{ScalarDM.PRL.2020.124.151301,ScalarDM.PRL2019.123.141102,ScalarDM.Urena-Lopez:2019kud}.

In this paper, we discuss the possibilities of searching dark matter candidates with the Compton process at the Shanghai laser electron gamma source (SLEGS) beamline.
First, the specifics of SLEGS will be given and following by showing the cross sections, as well as the differential cross sections of different DMPs generated in Compton processes. 
The possible experimental detection scheme will be discussed too. 
  
\begin{figure*}
\includegraphics[width=0.95\textwidth]{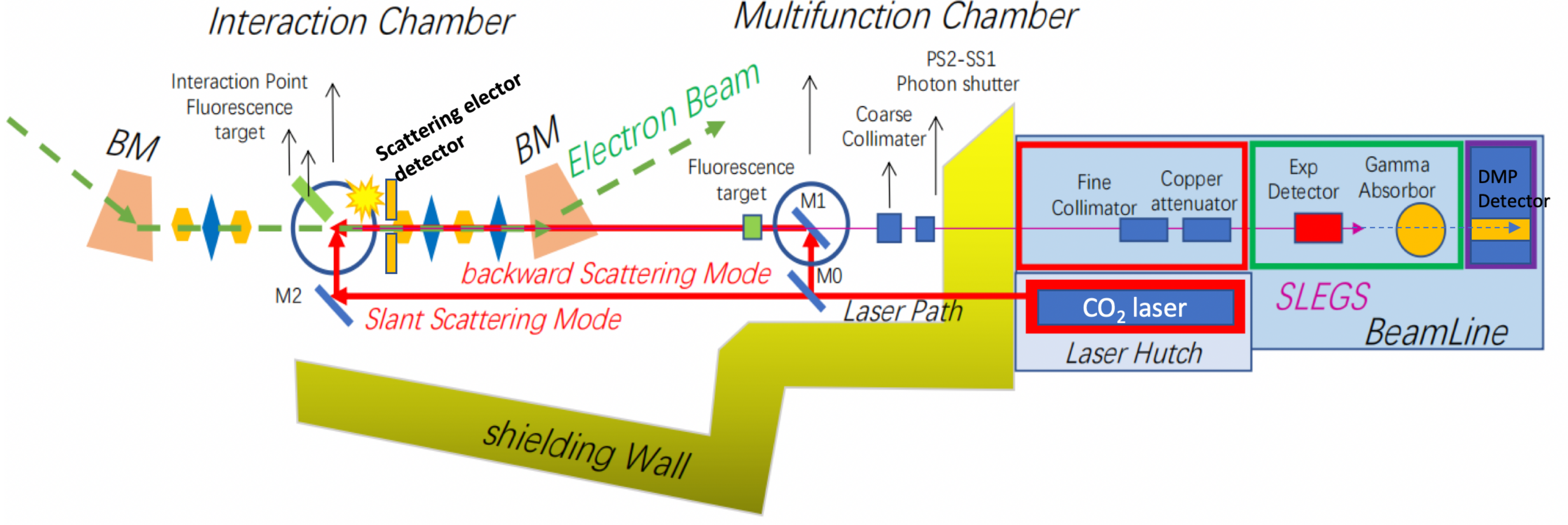}
\caption{  
Scheme of the SLEGS (not to scale).
3.5 GeV electron pulses move in the storage ring.
Photons from a CO$_2$ laser can be sent to collide with electrons in two ways:
one is in the line where the colliding angle is $180^\circ$,
and another way is by a tunable colliding angle between $20$-$160^\circ$.
Compton process products, including $\gamma$-ray and the possible DMPs,
are high concentred at zero degrees (which will be shown in details in the texts),
and will be detected by DMP detectors in the experimental area.
An electron detector is located around the target area for detecting scattering electrons.
It will serve as a start signal for TOF of the possible DMPs,
which will suppress background noise signals highly.
}
\label{fig.slegs}
\end{figure*}

\section{Shanghai laser electron gamma source beamline}

Shanghai laser electron gamma source (SLEGS) which is under construction is located at Shanghai Synchrotron Radiation Facility (SSRF), Shanghai, China\cite{SLEGS2008,SLEGS.LW.2010,SSRF.2009}.
It can produce a high intensity $\gamma$-ray beam by using laser Compton scattering (LCS) between  3.5 GeV electrons in the  storage ring of SSRF and photons from a CO$_2$ laser.
As shown in Fig.\ref{fig.slegs}, the 3.5 GeV electron beam moves circularly in the SSRF storage ring.
A polarized CO$_2$ laser beam ($\lambda= 10.64 \mu$m) is injected into the target area (interaction region) from the front-end downstream through the mirror system
and then collides with the electron beam head-on-head in the target chamber.
Through the LCS mechanism, $\gamma$-ray is generated in this process.
The LCS $\gamma$-rays are generated within a small forward cone along the moving direction of the incident electrons. 
The LCS $\gamma$ beam is then sent to the $\gamma$-ray experimental area through the thin reflecting mirror for CO$_2$ laser.
Collimators, degraders, as well as $\gamma$-ray beam diagnostic equipment, are  arranged in the way.

Feynman diagrams for the LCS in the lowest order are shown in Fig.\ref{fig.feyman3}.
In the SLEGS case, if a low energy photon from the CO$_2$ laser is absorbed, 
a higher energy  photon ($\gamma$-ray) will be emitted, considering electrons energy is 3.5 GeV.
The same as LCS processes, 
other rare processes related to DMP candidates may also take place when a photon collides with an electron. 
Instead of a real $\gamma$ photon, 
a pseudo scalar, scalar, or vector field particles, if exists, can be emitted, 
as shown in Fig.\ref{fig.feyman3}.
This process is called a Compton-like process.

\begin{figure}
\includegraphics[width=8.5cm]{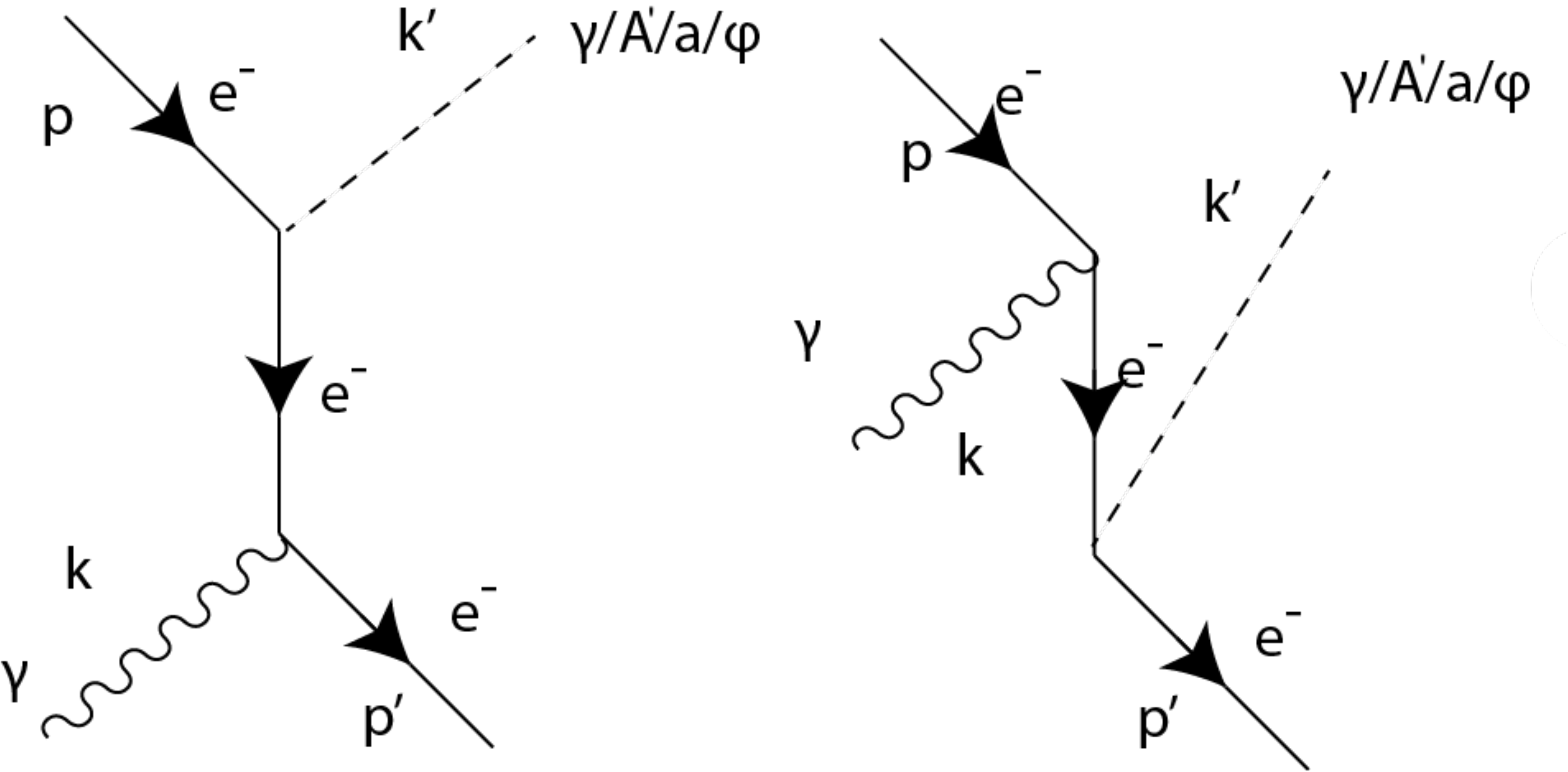}
\caption{  
Feynman Diagrams of Compton-like processes, 
which may produce a normal photon $\gamma$, or DMPs.
Here the DMPs could be dark photon $A'$,
axion $a$,
or scalar particle $\phi$.
They can be generated via this Compton-like process 
between a real photon and an electron\cite{DP-FILIPPI2020100042}.
}
\label{fig.feyman3}
\end{figure}

\section{Dark matter particles from Compton scattering processes}
Here we discuss different DMPs, including pseudo scalar, vector, and scalar particles, 
which may be generated in Compton processes.

\subsection{ Pseudo Scalars Particles}
If pseudo scalars particles  couple to electrons,
the corresponding effective Lagrangian can be written as\cite{EDELWEISS-Axion2013},
\begin{equation}
    \mathbf{L}= g_{ae}\frac{\partial_\mu \phi_a}{2m_e}\bar{\psi}_e \gamma^\mu \gamma^5 \psi_e=-i g_{ae}\bar{\psi}_e \gamma^5 \psi_e \phi_a,
\end{equation}
where dimensionless $g_{ae}$ is the pseudoscalars-electron coupling constant.
Here we consider a pseudo scalars particle which is emitted in a Compton-like process, i.e.
$\gamma+e^-\rightarrow e^- + a$.
The corresponding Feynman diagram is shown in Fig. \ref{fig.feyman3}.
The differential cross section can be written as,
\begin{eqnarray}\label{eq.dsigma.domega1}
\frac{d\sigma}{d\Omega}&=&\frac{1}{64\pi^2}\frac{1}{E_k E_p |v_k-v_p|}\frac{|\vec{k'}|^2}{k'^0 p'^0} \nonumber \\
&&\times\left\vert\frac{1}{ \frac{\vert\vec{k'}|}{ k'^0}+\frac{|\vec{k'} |-(\vec{p}+\vec{k})\cdot \hat{k'}}{p'^0}}\right\vert\left(\frac{1}{2}\sum_{spin} |M^2|\right),
\end{eqnarray}
where the definitions of  $\vec{k}$, $\vec{k}'$,  $\vec{p}$, $v_k$, $v_p$, $E_k$, $E_p$, and $M$ etc. can be found in the appendix. $spin$ is the incoming electron's spin. 

Then the total cross section in laboratory coordinator can be obtained by integral,
\begin{eqnarray}\label{eq.CS.lab}
\sigma_{lab}&=&\int d\sigma 
= \prod_f \Big(\int  \frac{d^3 \vec{p}_f}{(2\pi)^3} \frac{1}{2E_f}\Big)   \frac{1}{2E_k 2E_p |v_k-v_p|} \nonumber \\
&&\times(\frac{1}{2}\sum_{spin} |M^2|)(2\pi)^4 \delta^4 (p+k-p'-k'),
\end{eqnarray}
 where subscript index $f$ represents the final quantum state of dark matter candidate (pseudo scalar) and electron.

From the Eq.\ref{eq.dsigma.domega1}, 
one can see that 
the differential cross section $\frac{{\rm d}\sigma_{lab}}{{\rm d}\Omega}(m_a, \theta_0, \theta, \phi)$ 
is a function of the DMP's mass $m_a$, 
colliding angle between the electron and the laser $\theta_0$,
and scattering angles, $\theta$ and $\phi$.
While the total cross section $\sigma_{lab}(m_a, \theta_0)$  depends on $m_a$ and $\theta_0$ only.

By using the parameters of SLEGS, 
$E_e=3.5$ GeV and photon's wavelength 10.64 $\mu$m, 
the cross section $\sigma_{lab}(m_a, \theta_0)$ 
is shown in Fig.\ref{fig.CS.pScalar},
where 4 colliding angles, $\theta_0=90^\circ$, 
$120^\circ$,
$150^\circ$,
and $180^\circ$,
are drawn.
As one can find in the figure, 
at $\theta_0=180^\circ$, i.e. head-to-head collision,
the cross section approaches its maximum.
It can also be seen that  the cross section is decreasing while the pesudoscalar's mass increasing.
The differential cross section $\frac{{\rm d}\sigma_{lab}}{{\rm d}\Omega}(m_a, \theta_0, \theta, \phi)$ can be found in the appendix.

\begin{figure}
\centering
\includegraphics[width=9cm]{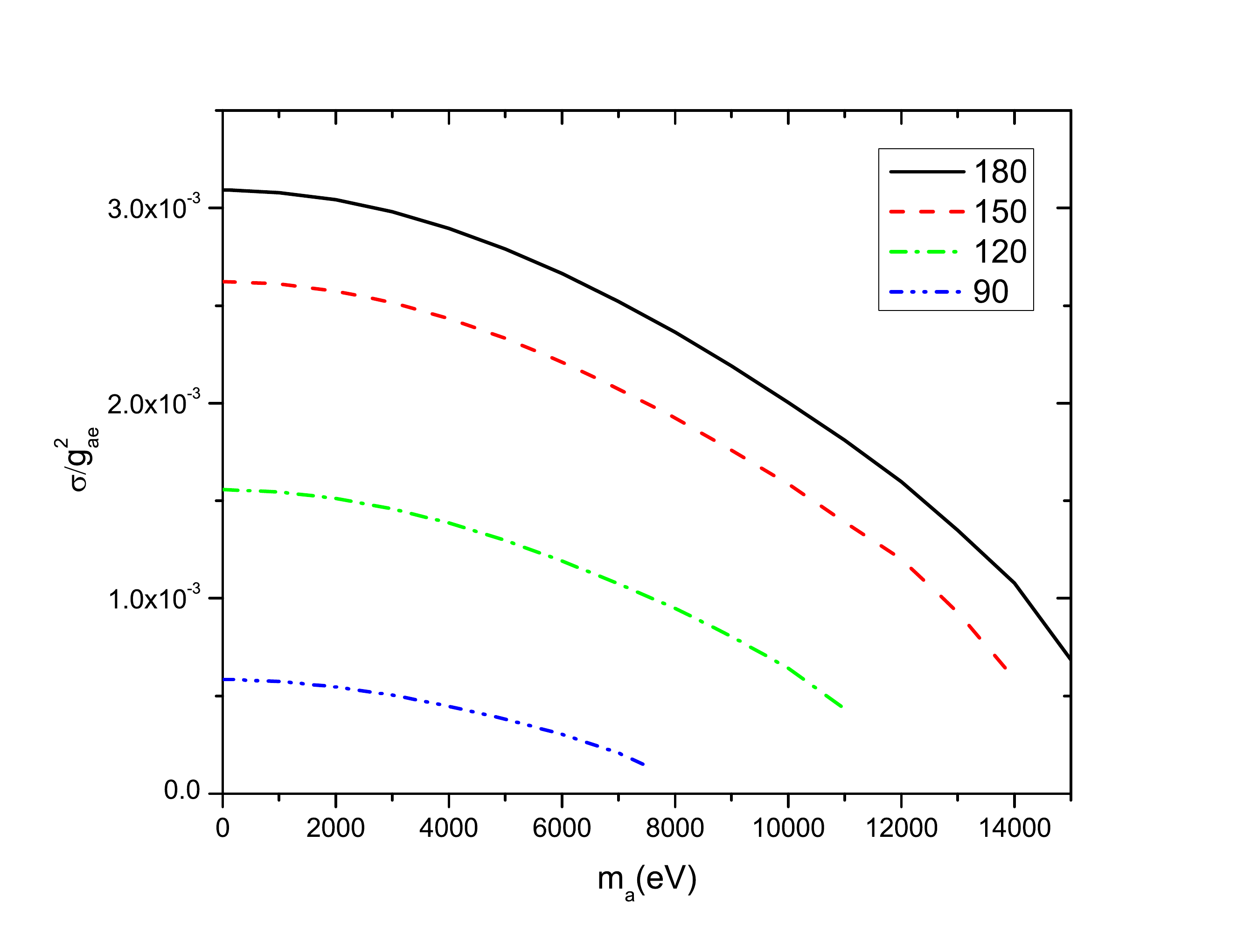}

\caption{  
Pseudo scalar generating cross sections through reaction $e+\gamma \rightarrow e+a$.
The cross section  $\sigma_{lab}(m_a, \theta_0)$ is a function of the pseudo scalar's mass $m_a$, 
and colliding angle $\theta_0$ between the pseudo scalar and electron.
The different color lines represent $\theta_0=90^\circ$ (blue dash dot dot), 
$120^\circ$(green dash dot),
$150^\circ$(red dash),
and $180^\circ$(black solid). The cross section $\sigma$ is in unit of barn.
}
\label{fig.CS.pScalar}
\end{figure}

\subsection{Dark Photon}

For vector field particles, like dark photons, coupling to an electron case,
the corresponding effective Lagrangian can be written as\cite{DMP-Chakrabarty:2019kdd},
\begin{equation}
    \mathbf{L}=  g_{A'e} e \bar{\psi}_e \gamma^\mu \psi_e A'_\mu,
\end{equation}
where dimensionless $g_{A'e} $ is the mixing parameter between the dark photon and the Standard Model photon, and $e$ is the electron charge.
The same as the Compton process, 
the dark-photon-electron interaction process can be written as,
$\gamma+e^-\rightarrow e^- + A'$.
The corresponding Feynman diagram is shown in Fig. \ref{fig.feyman3}.

The interaction's  differential cross section and total cross section are described in details in the appendix. 
The same as the pseudo scalar case, 
 taking electron energy $E_e=3.5$ GeV and the laser polarization direction perpendicular to the reaction plane,
the cross section $\sigma_{lab}(m_a, \theta_0)$ for interaction $\gamma+e^-\rightarrow e^- + A'$
is shown in Fig.\ref{fig.CS.DarkPhothon}.
Here, $\sigma_{lab}(m_a, \theta_0)$  with colliding angles, $\theta_0=90^\circ$, 
$120^\circ$,
$150^\circ$,
and $180^\circ$,
are shown in drawn.
Again, head-to-head collision with  $\theta_0=180^\circ$
has the maximum cross section compared with other collision angles.
The cross section for the dark photon cases drops too with increasing of the dark photon's mass.
The dependence of  the differential cross section on scattering angles in dark photon case  is shown in the appendix.

\begin{figure}
\centering
\includegraphics[width=9cm]{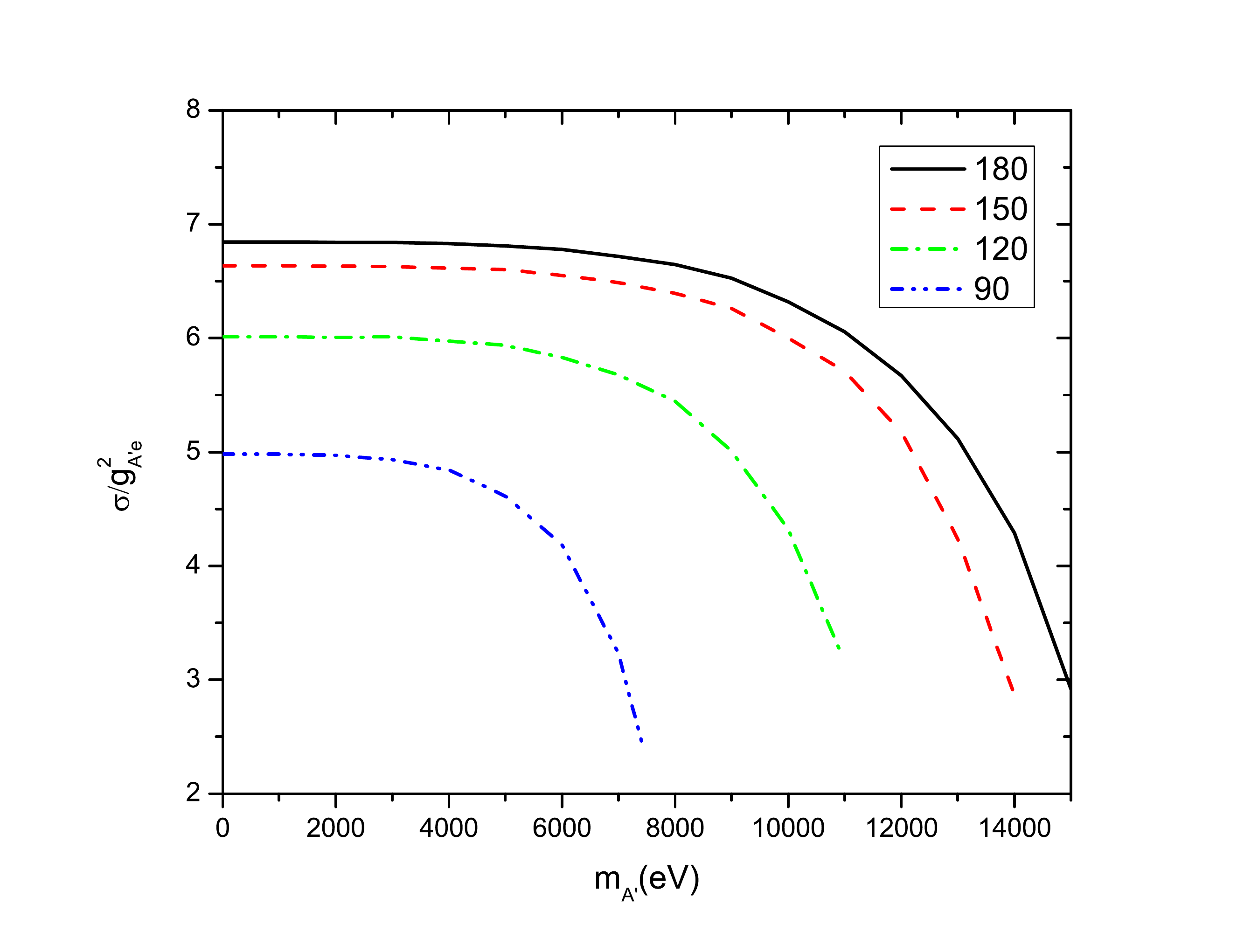}

\caption{  
Dark photon generating cross sections through reaction $e+\gamma \rightarrow e+\phi$.
The cross section  $\sigma_{lab}(m_a, \theta_0)$ is a function of dark photon's mass $m_{A'}$, 
and colliding angle $\theta_0$ between dark photon and electron.
The different color lines represent $\theta_0=90^\circ$ (blue dash dot dot), 
$120^\circ$(green dash dot),
$150^\circ$(red dash),
and $180^\circ$(black solid). The cross section $\sigma$ is in unit of barn.
}
\label{fig.CS.DarkPhothon}
\end{figure}

\subsection{Scalar Dark Matter Particle}

For scalar dark matter particle case, $\gamma+e^-\rightarrow e^- + \phi$,
the corresponding effective Lagrangian can be written as\cite{ScalarDM-PhysRevD.96.115021},

\begin{equation}
   	L=g_{\phi e} \bar{\psi}_e \psi_e \phi,
\end{equation}
where dimensionless $g_{\phi e}$ is the scalar-field-partile-electron coupling constant.
The corresponding Feynman diagram is shown in Fig. \ref{fig.CS.scalar}.

Taking electron energy $E_e=3.5$ GeV,
the cross section $\sigma_{lab}(m_a, \theta_0)$ for interaction $\gamma+e^-\rightarrow e^- + \phi$
can be calculated, and is shown in Fig.\ref{fig.CS.scalar}, with for scattering angles, $\theta_0=90^\circ$, 
$120^\circ$,
$150^\circ$,
and $180^\circ$.
For the same mass of a scalar DMP, the interaction cross section increases with scattering angle increasing,
and reaches to its maximum  at $\theta_0=180^\circ$.
Detecting at a given angle, the expected cross section drops with increasing of the scalar DMP's mass.

Further information about the interaction's  differential cross section and total cross section are described  in the appendix. 
As one can see from there, the cross sections have different dependence on the scattering angles. In the calculations, the polarization of the laser beam is parallel to $y$-axis. The cross section of pseudo scalar particles is insensitive to the incoming photon’s polarization (Fig. \ref{fig.DCS.pScalar}). However, the cross sections of dark photons and scalar particles are highly dependent on the laser beam’s polarization (Fig.\ref{fig.DCS.DarkPhothon}\&\ref{fig.DCS.scalar}). The dark photons are mainly emitted parallel to the $x$-direction, which is similar to the Thomson scattering. On the other hand, the emission of the scalar particles is mainly parallel to the $y$-direction (Fig. \ref{fig.DCS.scalar}). It is because the radiation of scalar particles is parallel to the laser’s polarization. The angular distributions can be used to distinguish different types of dark matter candidates since the distributions highly depend on particle types.

\begin{figure}
\centering
\includegraphics[width=9cm]{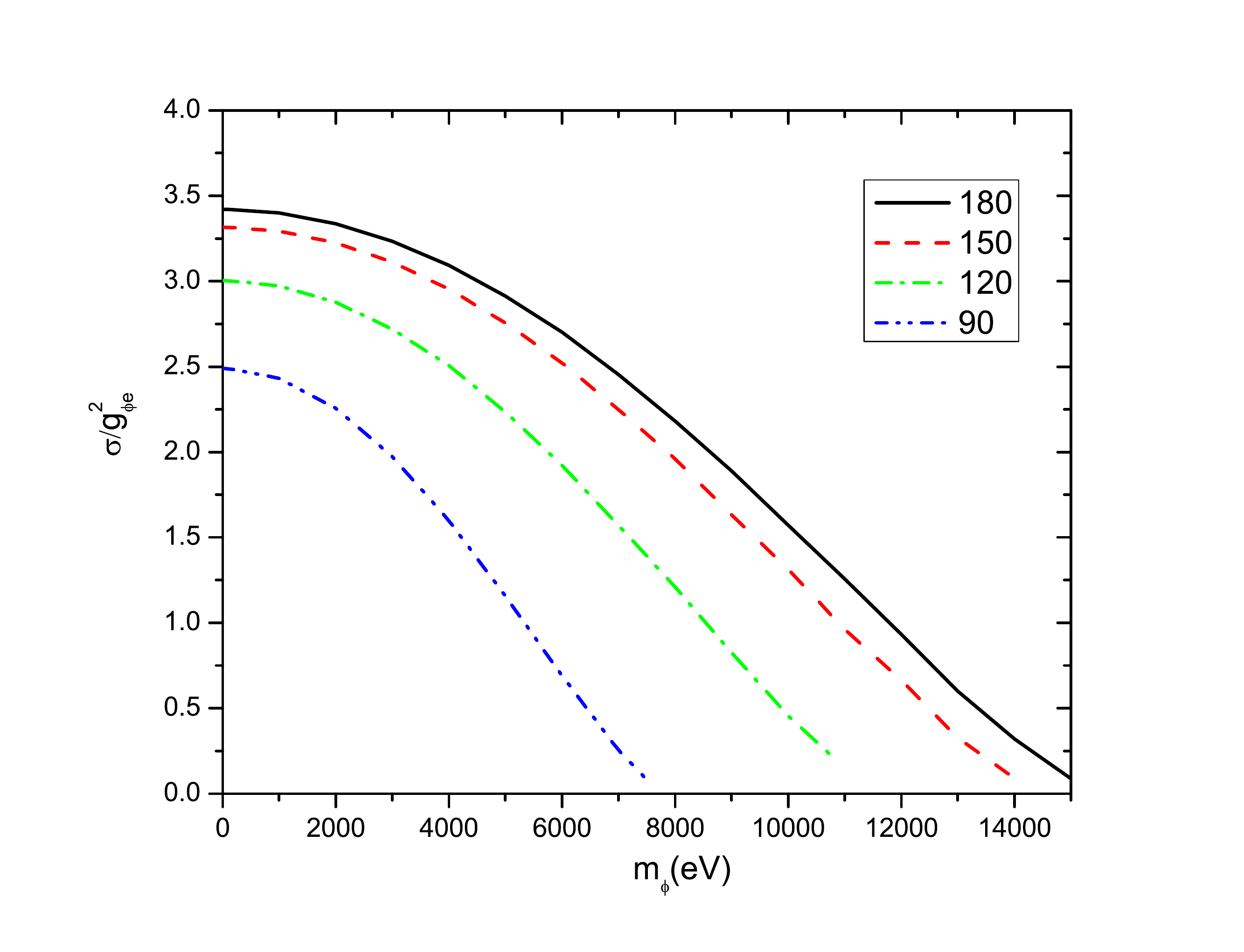}
\caption{  
Scalar generating cross sections through reaction $e+\gamma \rightarrow e+\phi$.
The cross section  $\sigma_{lab}(m_a, \theta_0)$ is a function of scalar DMP's mass $m_\phi$, 
and colliding angle $\theta_0$ between the scalar DMP and electron.
The different color lines represent 
$180^\circ$(black solid),
$150^\circ$(red dash),
$120^\circ$(green dash dot),
and $\theta_0=90^\circ$ (blue dash dot dot). The cross section $\sigma$ is in unit of barn.
}
\label{fig.CS.scalar}
\end{figure}

\section{ Dark Matter Particle Searching with Electron Compton Scattering}

 \begin{figure}
\includegraphics[width=0.48\textwidth]{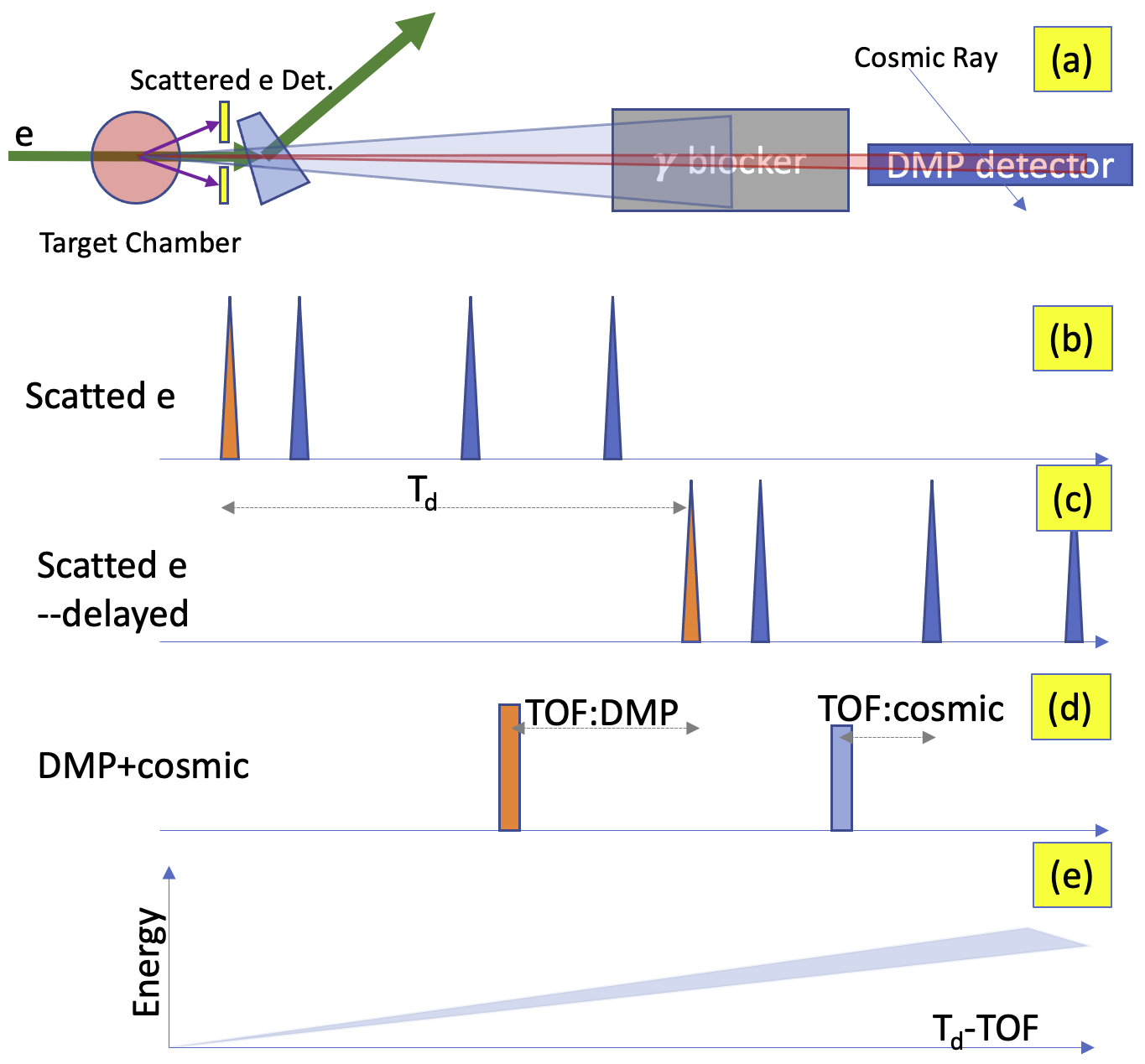}
\caption{
Inverse time-of-flight (TOF) method for the possible DMPs. 
(a) Detecting scheme for the possible DMPs. 
The $\gamma$ photons are stopped by the $\gamma$-blocker;
The possible DMPs are detected by the DMP detector through reaction $X+e\rightarrow e+\gamma$;
The scatted electrons from the inverse Compton scattering are detected by a detector array.
(b) Signals ($S_e$) recorded by the electron detector array. 
(c) Delayed signals ($S^(d)_e$) of ($S_e$). The delay time is $T_d$.
(d) Signals from the DMP detectors, including these from the possible DMPs, cosmic rays, and/or other ambient radiations.
(e) Energy vs. $(T_d-TOF)$ spectrum, where shaded area represents signals from DMPs which have relatively high energies (MeV level) and low rest mass (keV level).
}
\label{fig.bkgd}
\end{figure}
 \begin{figure}[ht]
\includegraphics[width=0.48\textwidth]{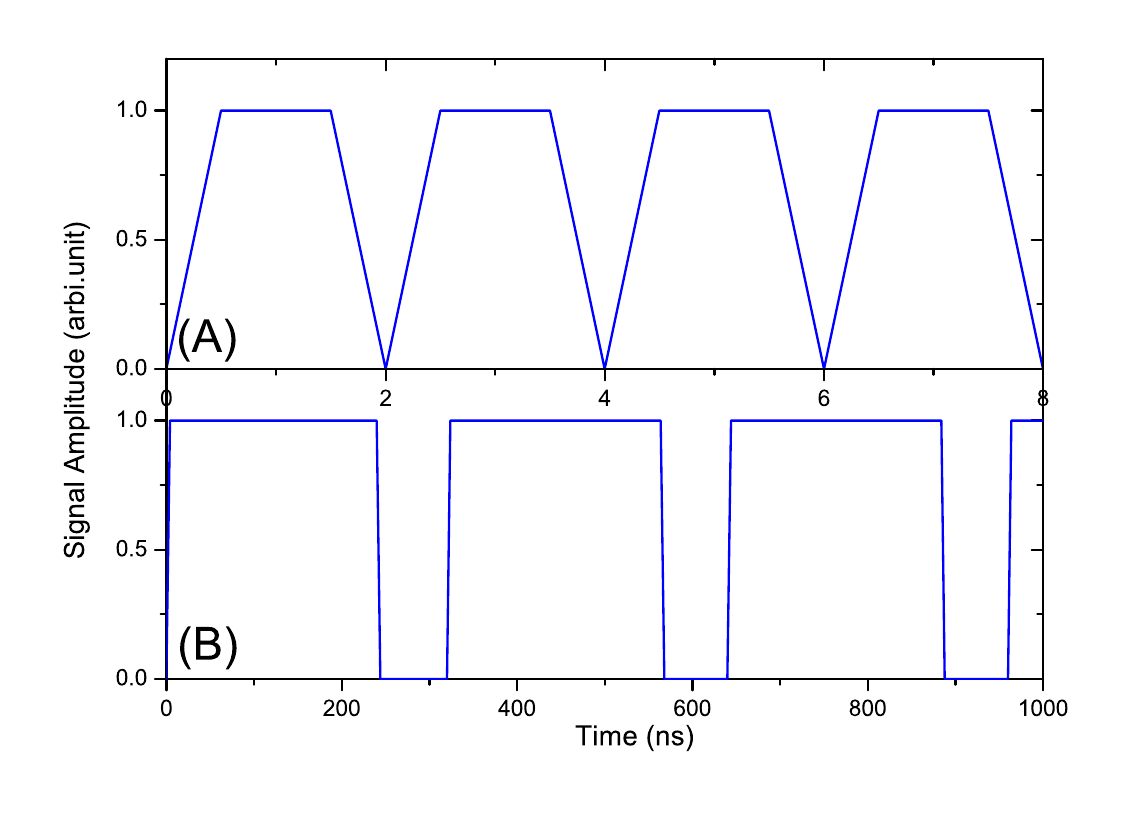}
\caption{  
Schematic diagram of a typical electron beam structure at SSRF.
The electron beam is composed of small bunches, which have widths smaller  than 1 ns\cite{SSRF.2009}.
multiple bunches can be combined to be a macro pulse.
(A) is the zoomed-in structure of the beam in time range 0 to 8 ns;
and (B) is the zoomed-out structure in range 0 to 1000 ns.
A macro pulse that has 120 single bunches inside has a width of 240 ns.
}
\label{fig.time.beam}
\end{figure}

DMPs may be searched by putting a detector set at the experimental area and an electron detector set near the target, as 
shown in Fig.\ref{fig.bkgd} (also in Fig.\ref{fig.slegs}). 
After the laser-electron collision,
scattered electrons can be detected by  the electron detector set,
while the $\gamma$-photons, or the DMPs, will be emitted in a narrow forward solid angle (see Fig.\ref{fig.DCS.pScalar}-\ref{fig.DCS.scalar} in the appendix). 
The $\gamma$-photons will be absorbed by the $\gamma$-blocker,
and the possible DMPs can pass through the $\gamma$-blocker can be recorded by the DMP detector through the reaction $X+e\rightarrow e+\gamma$.

Because the rate of the scatted electron is much larger than that of DMP,
an inverse-TOF method could be used to get the TOF of the possible DMPs,
as shown in the Fig.\ref{fig.bkgd}b-d.
A  signal from the DMP detector is used as the ``start'' of the TOF, 
while delayed signals from the electron detector set are  used as the ``stop''.
With this method, the rate of the TOF of DMP is highly reduced, and then results in a high reduction of the random coincidence rate.
In the energy vs. TOF spectrum, to-be-discovered DMP signals are located in a very narrow area,
because of their high energy (MeV level) and low rest mass (keV level).
The DMP detector can also record the background signals including the ambient radiation and cosmic rays. 
Because the background signals come in randomly, 
this kind of data is randomly distributing in the TOF vs energy spectrum, 
and only a very few of them filling in the region of interest.
Furthermore, the cosmic rays and ambient radiations can be further reduced with  an anti-cosmic-muon device and passive shielding. 

The time and energy resolutions of the DMP detector determine the rejection rate for the random background signals.  
With modern detecting technologies, 
time-of-flight (TOF) precision of about tens of ps can be achievable,
as well as energy resolution of less than 1\%.

The inverse-TOF method can only work with a well-defined electron beam structure. 
A typical electron beam structure is shown in Fig.\ref{fig.time.beam}.
The linac can operate in two modes:
single bunch mode or multi-bunch mode\cite{SSRF.2009}.
Each bunch has a width of  less than ns. 
The time interval between two bunches is 2 ns.
In multi-bunch mode,  many single bunches can be combined to be a macro pulse.
For example, as shown in Fig.\ref{fig.time.beam},  
a macro pulse which has 120 bunches inside has a width of 240 ns,
following by an 80 ns (40 periods) silence time, 
and then repeating the pattern again and again.
Therefore, the well-defined electron beam structure provides a good start point for the DMP data analysis.

With assumptions of a continuous-wave laser with a power of 1000 kW,
a laser focus diameter of 2 $\mu$m,
the electron beam energy of 3 GeV,
the electron beam intensity of 240 mA,
the time resolution of the TOF 0.1 ns,
the detector length of 30 m,
and two years of data collecting time, 
the DMPs detecting rates have been simulated.
The results of the constraints on  pseudo scalar, dark photon, and scalar particles are shown in Figs. \ref{fig.limit.ALP}-\ref{fig.limit.ScalarDM}. 

To further improve the constraints on the DMP coupling constants, a pulsed laser could be used.
The peak power of a pulsed laser could be in the range of  several GW to even TW with today's technologies.
With the increasing of the laser's peak power $P_{L}$,
the constraints on the DMP coupling constants  can be reduced by $\propto P_{L}^{1/2}$. 
By adding an optical cavity, there has a potential to improve the constraints by a factor over 1000\cite{Cavity_Couprie_1999, Cavity_Bonis_2012}.

There has another advantage.
When SLEGS running, it does not interfere with other experiments which are running at the SSRF.
Furthermore, when DMP experiments running, it does not interfere $\gamma$-ray experiments which are running on SLEGS too.
A DMP experiment will just keep quiet.
Whenever the facility is running, it records data.

There has  another electron facility, Shanghai HIgh repetitioN rate XFEL aNd Extreme light facility (SHINE)\cite{SHINE-PhysRevAccelBeams.22.090701},  under construction in the Shanghai area. It will have energy up to 8.8 GeV, 100 pC/bunch, and  a repeating frequency of 1 MHz, 
which makes it possible for the DMP searching experiments.

\begin{figure}
\centering
\includegraphics[width=8cm]{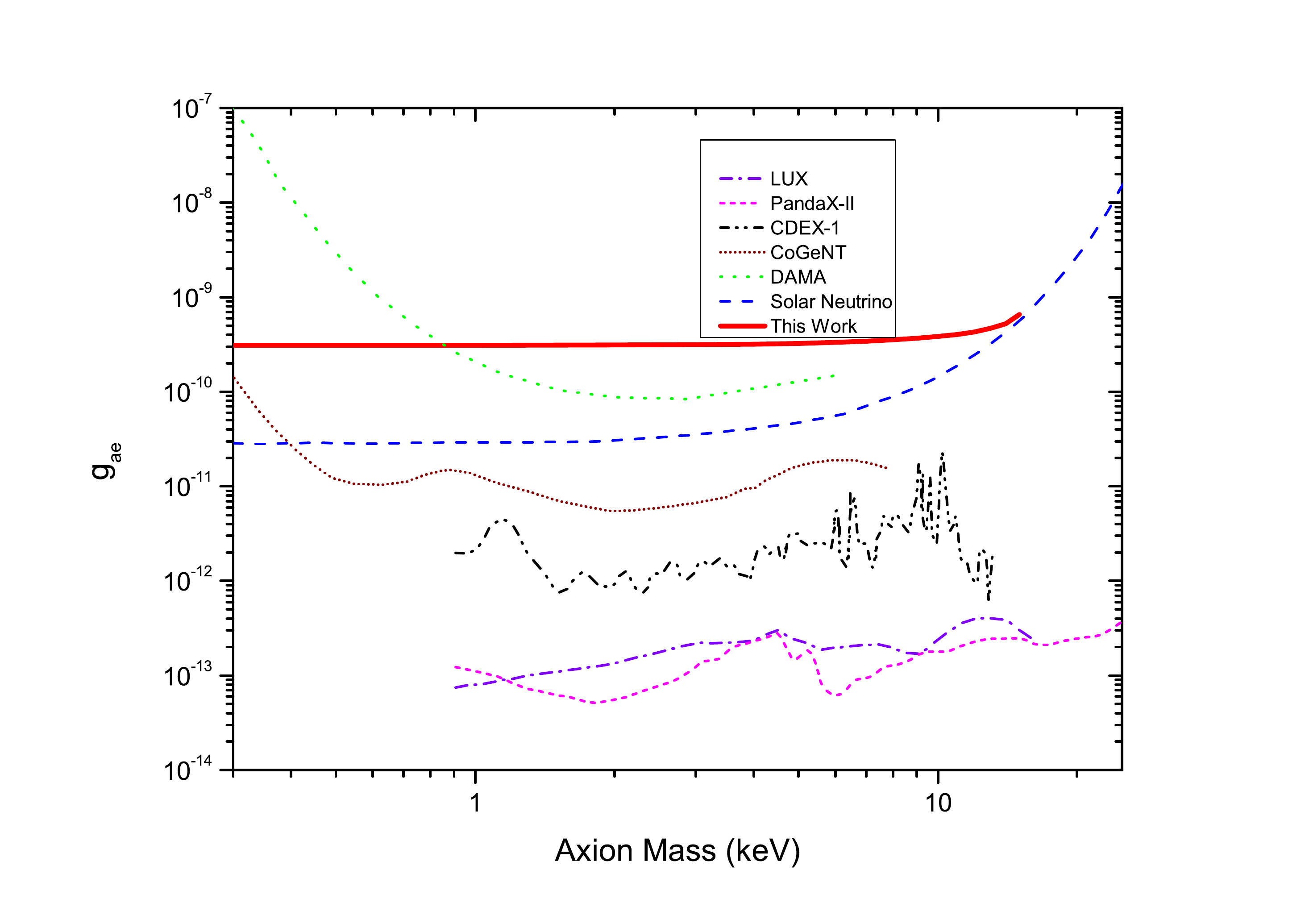}
\caption{  
The pseudoscalar-electron coupling limits.
The lines in the figure represent limits from
LUX (purple dash dot)\cite{LUX2017.PhysRevLett.118.261301},
PandaX-II (pink dash)\cite{PandaX2017PhysRevLett.119.181806},
CDEX-1(black dash dot dot)\cite{CDEX2017.PhysRevD.95.052006},
CoGeNT(brown dot)\cite{CoGeNT2011.PhysRevLett.106.131301},
DAMA(green short dash)\cite{DAMA2006RN93},
solar neutrino\cite{SolarNeutrino2009PhysRevD.79.107301}(blue dash),
and this work (red solid) with parameters of a pulse laser with peak power 1000kW and 2 years data collecting.
}
\label{fig.limit.ALP}
\end{figure}

\begin{figure}
\centering
\includegraphics[width=8cm]{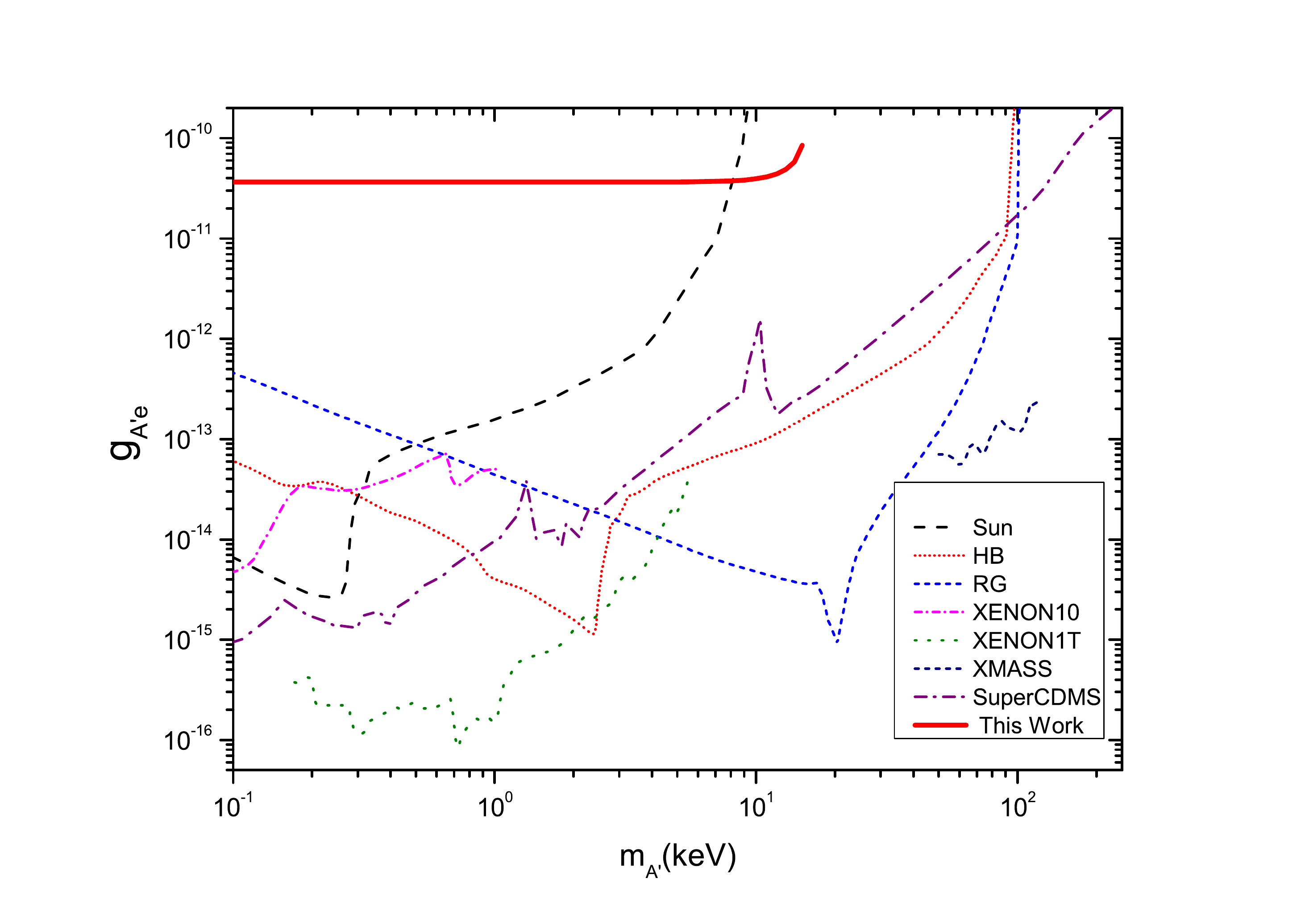}
\caption{  
The electron-dark-photon coupling limits.
The lines in the figure represent limits from
the Sun (black dash)\cite{Xn1T2017.RN97},
horizontal branch stars (HB, pink dot)\cite{Xn1T2017.RN97},
red giant (RG, blue short dash)\cite{Xn1T2017.RN97},
XENON10(pink dash dot) and XENON1T (green dot)\cite{Xn1T2017.RN97},
XMASS (black short dash)\cite{XMass2020.RN95},
SuperCDMS (purple dash dot)\cite{SuperCDMS2020-PhysRevD.101.052008},
and this work (red solid) with parameters of a pulse laser with peak power 1000 kW and 2 years data collecting.
}
\label{fig.limit.DP}
\end{figure}

\begin{figure}
\centering
\includegraphics[width=8cm]{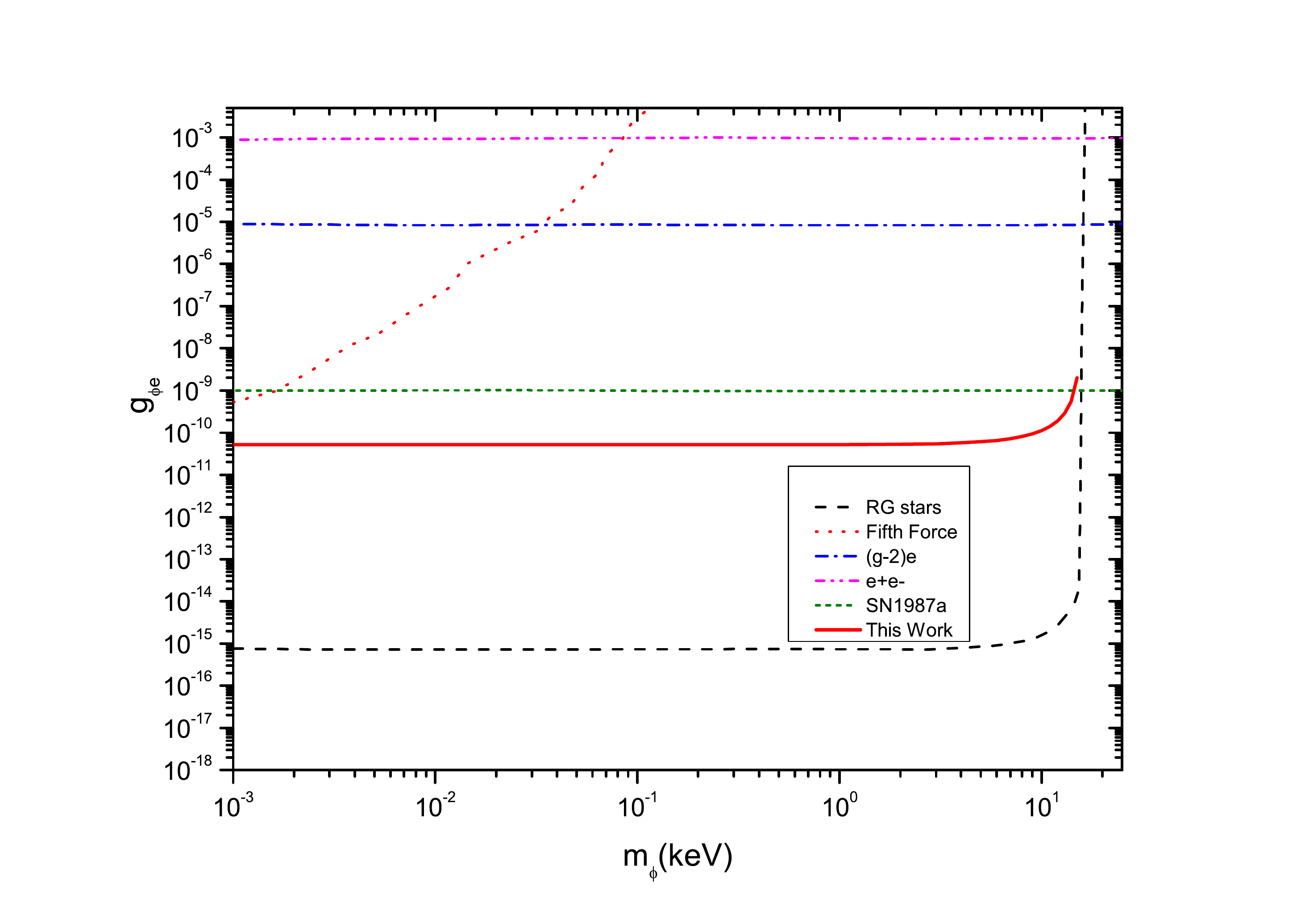}
\caption{  
The electron-scalar-DMP coupling limits.
The x-axis $m_\phi$ represents the rest mass of the scalar DMP.
The lines in the figure represent limits from
red giant stars (black dash) \cite{RG-star-RN94,LightDM-PhysRevD.96.115021},
fifth force searches (orange dot)\cite{5F-Murata:2014nra,LightDM-PhysRevD.96.115021},
electron g-2 (blue dash dot)\cite{g-2e-PhysRevLett.117.101801,LightDM-PhysRevD.96.115021},
$\phi\rightarrow e^+e^-$ (pink dash dot dot)\cite{g-2e-PhysRevLett.117.101801,LightDM-PhysRevD.96.115021},
supernova SN1987a (green short dash)\cite{RG-star-RN94},
and this work (red solid) with parameters of a pulse laser with peak power 1000 kW and 2 years data collecting.
}
\label{fig.limit.ScalarDM}
\end{figure}

\section{summary}
In summary, the possibilities of searching for dark matter particles by using the SLEGS beamline are discussed. 
One advantage of using electron-photon Compton scattering to search for the DMPs is that the scattered DMPs are highly concentrated in forwarding  angles, which makes the detecting easier.
By using electron detectors as start signals of TOF, 
the background noise, which is the key issue in extremely rare event detecting, can be highly suppressed.
It is promising that  electron facilities could be new platforms for  searching light dark matter particles like axions, dark photons, or dark scalar particles. 

\begin{acknowledgments}
This work is supported by
the National Nature Science Foundation of China (grant Nos. 11875191 and 11775140),
and the Strategic Priority Research Program (grant No. CAS XDB1602).
\end{acknowledgments}


\section{Appendix}

The scattering cross section can be obtained directly from the S-matrix.  

\begin{eqnarray}
\frac{d\sigma}{d\Omega}&=&\frac{1}{64\pi^2}\frac{1}{E_k E_p |v_k-v_p|}\frac{|\vec{k'}|^2}{k'^0 p'^0}
\times\left\vert\frac{1}{ \frac{\vert\vec{k'}|}{ k'^0}+\frac{|\vec{k'} |-(\vec{p}+\vec{k})\cdot \hat{k'}}{p'^0}}\right\vert\left(\frac{1}{2}\sum |M^2|\right), \\
d\sigma &=&\frac{1}{2E_k 2E_p |v_k-v_p|} \left( \prod_f \frac{d^3 p_f}{(2\pi)^3} \frac{1}{2E_f} \right)(\frac{1}{2}\sum_{spin} |M^2|)(2\pi)^4 \delta^4 (p+k-p'-k'),
\end{eqnarray}
where we replace $|M|^2$ with $\frac{1}{2}\sum_{spin}|M|^2$, because the electron's spin is not controllable. It is the average of all possible electrons' spins.   
z-axis is chosen to be the moving direction of the incoming electron. The photon and electron are located at x-z plane.  The four momenta of all particles in the Cartesian coordinates are  
$p=(p^0,0,0,|\vec{p}|)$, 
$k=(k^0,|\vec{k}|\sin\theta_p,0,|\vec{k}|\cos\theta_p)$,
and $k'=(k'^0, |\vec{k'})|\sin\theta \cos\phi, |\vec{k'})|\sin\theta \sin\phi|,\vec{k'})|\cos\theta )$.  The four momenta of emitted particles, $p'$ and $k'$, can be found by solving the energy momentum conservation and energy momentum relation, $p^2=p'^2= m_e^2$ and $k'^2=m_A^2$.

For different types of particles, 
pseudo scalar, scalar, or vector, 
the $\frac{1}{2}\sum_{spin}|M|^2$ can be written as in following.
\subsection{pseudo scalar}

There are two possible Feynman diagram to contribution $e^- \gamma \rightarrow e^- \psi_a$, fig. \ref{fig.feyman3}.  
In pseudo scalar case, the $\frac{1}{2}\sum_{spin}|M|^2$ is,
\begin{eqnarray}
\frac{1}{2}\sum_{spin}|M|^2&=&-g_{ae}^2 e^2 \left(\frac{I}{\Big((p+k)^2-m_e^2\Big)^2}     
+\frac{II}{\Big((p+k)^2-m_e^2\Big) \Big((p-k')^2-m_e^2\Big)}
+\frac{III}{\Big((p-k')^2-m_e^2\Big)^2}\right),
\end{eqnarray}
The nominators are    
\begin{eqnarray}
I&=&4(p'\cdot k)(k \cdot p)+8( p'\cdot k)( p \cdot \epsilon) (p \cdot \epsilon)-8( p'\cdot \epsilon)( k\cdot p)(p \cdot \epsilon)+8( p'\cdot p)( p\cdot \epsilon)( p\cdot \epsilon)-8m_e^2 ( p\cdot \epsilon)( p\cdot \epsilon)\\
\frac{II}{2}&=&4(p \cdot k)(p' \cdot k)+4( p'\cdot \epsilon)( p'\cdot k)(p \cdot \epsilon )-4(p' \cdot \epsilon )(p' \cdot \epsilon)( p\cdot k)+4(p \cdot \epsilon)( p'\cdot k)( p\cdot \epsilon)-4( p\cdot \epsilon)(p' \cdot \epsilon)( p\cdot k)\\
&&+8( p\cdot \epsilon)(p' \cdot \epsilon)(p' \cdot p)-8m_e^2 ( p\cdot\epsilon )( p'\cdot \epsilon)\\
III&=&4(p \cdot k)(p' \cdot k)-8(p' \cdot \epsilon )(p' \cdot \epsilon )( k\cdot p)+8( p'\cdot \epsilon)( p'\cdot k)( p\cdot \epsilon)+8( p'\cdot p)(p' \cdot \epsilon)(p' \cdot \epsilon)-8m_e^2 ( p'\cdot \epsilon)( p'\cdot \epsilon)
\end{eqnarray} 
where $\epsilon$ is the polarization direction of the photon. 
In this case, it points to the y-direction. 
The differential cross section at different collision angle 
$\theta=90^\circ, 120^\circ, 150^\circ$, and $180^\circ$ are shown in Fig.\ref{fig.DCS.pScalar}. 
As expected, because the energy of the electron is much larger than that of the photon, 
in the lab frame, the expected scalar DMPs are highly concentrated in the forward angle.
This property will benefit experimental detection.
\begin{figure*}
\centering
\subfigure{
\includegraphics[width=0.45\textwidth]{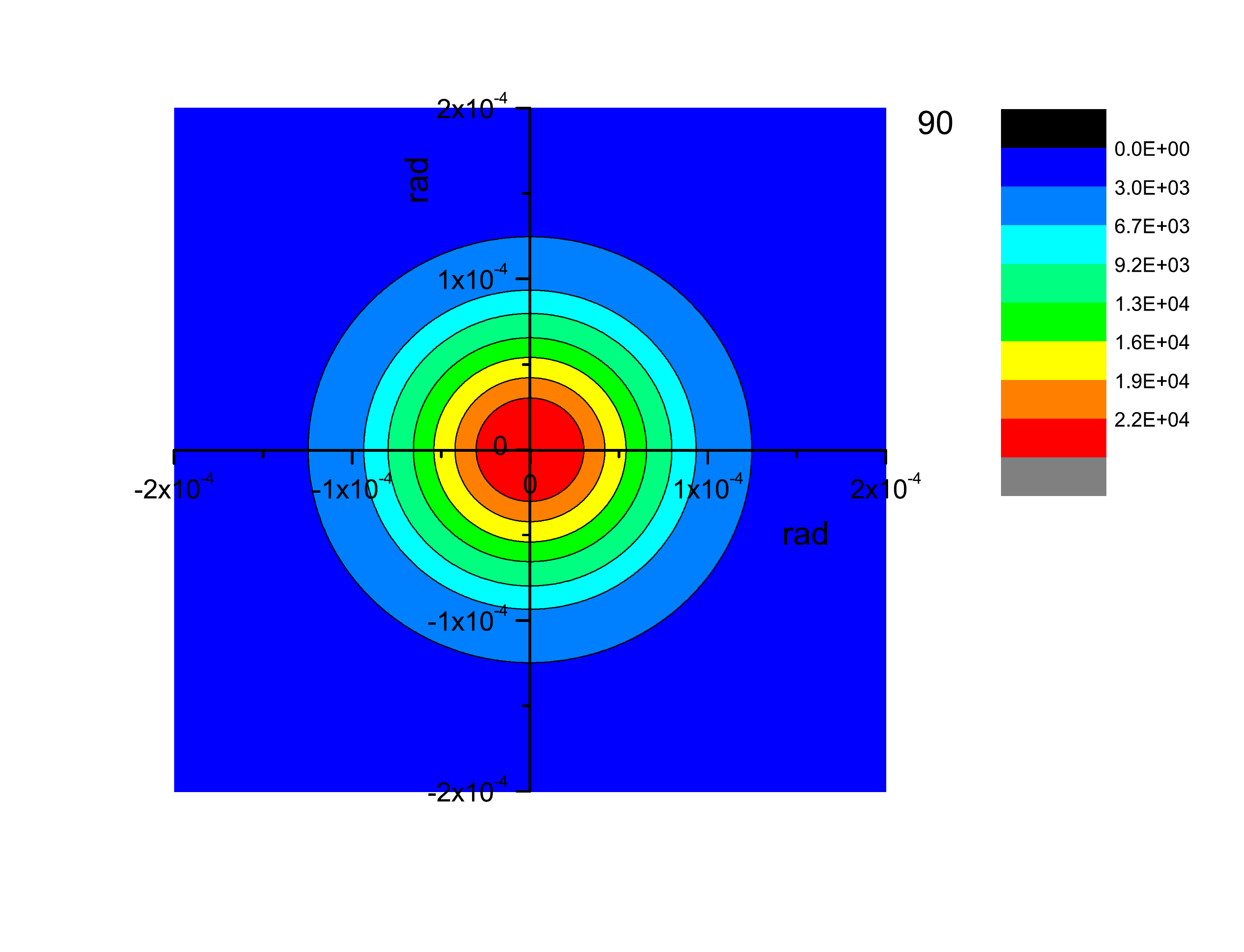} 
\includegraphics[width=0.45\textwidth]{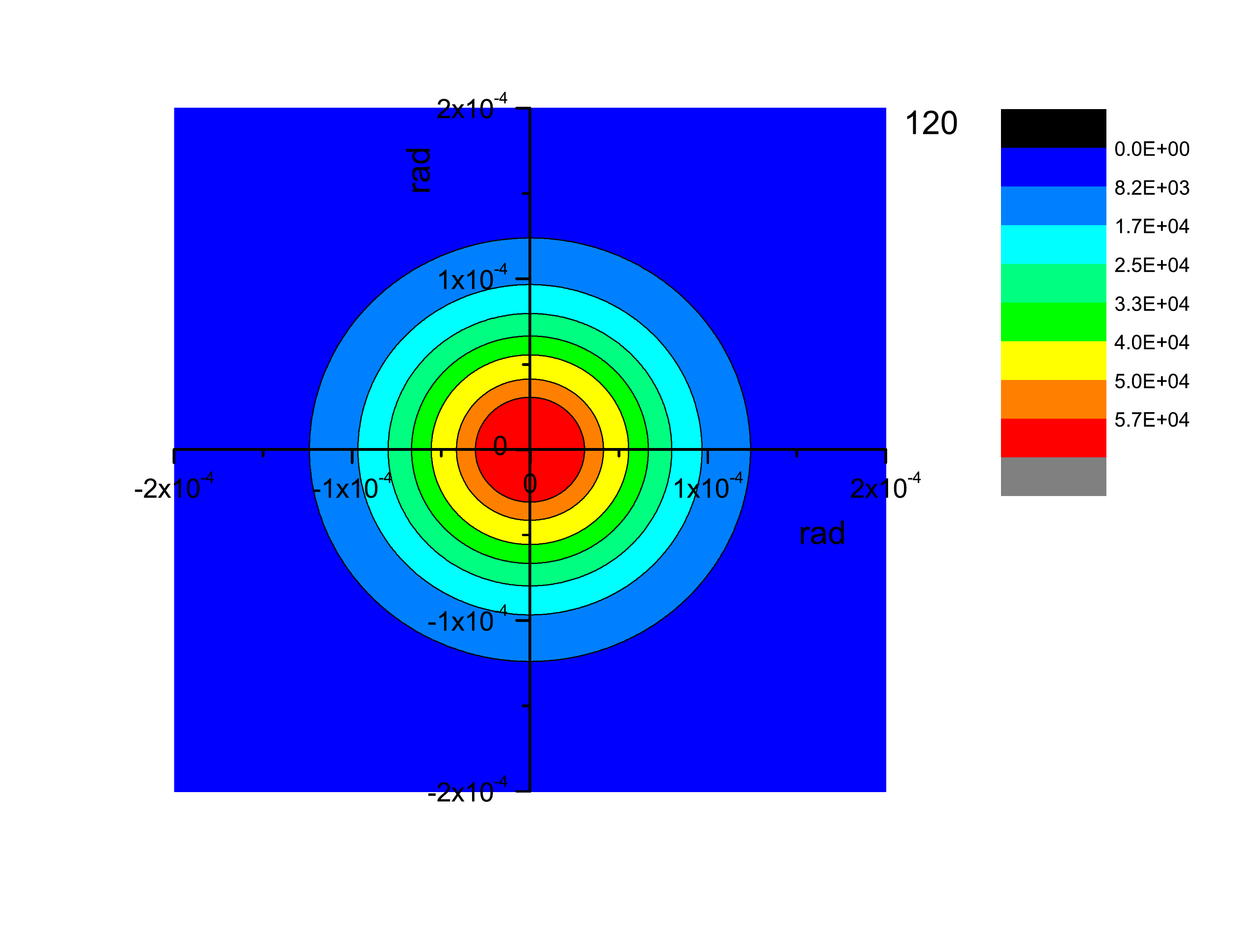} 
}
\subfigure{
\includegraphics[width=0.45\textwidth]{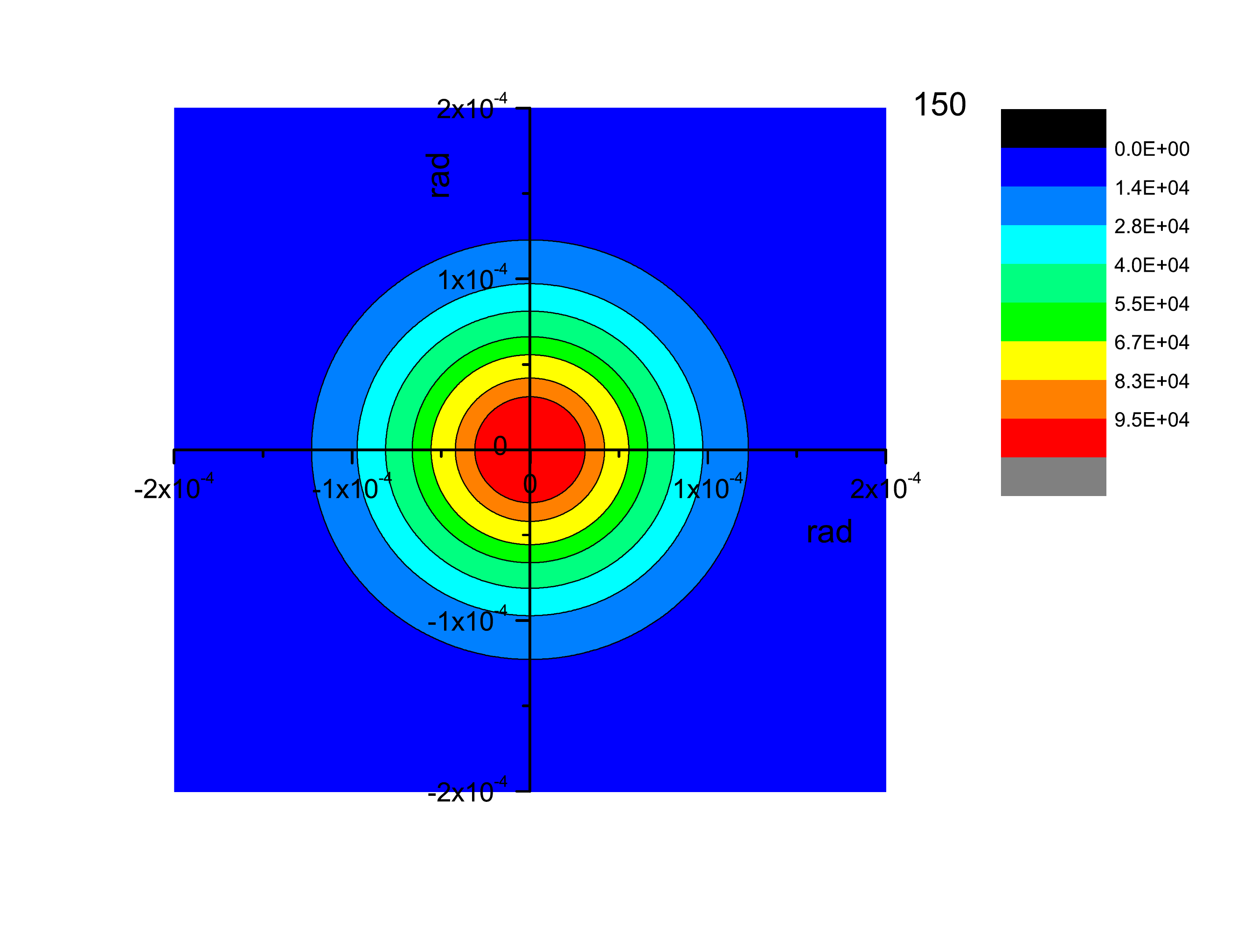} 
\includegraphics[width=0.45\textwidth]{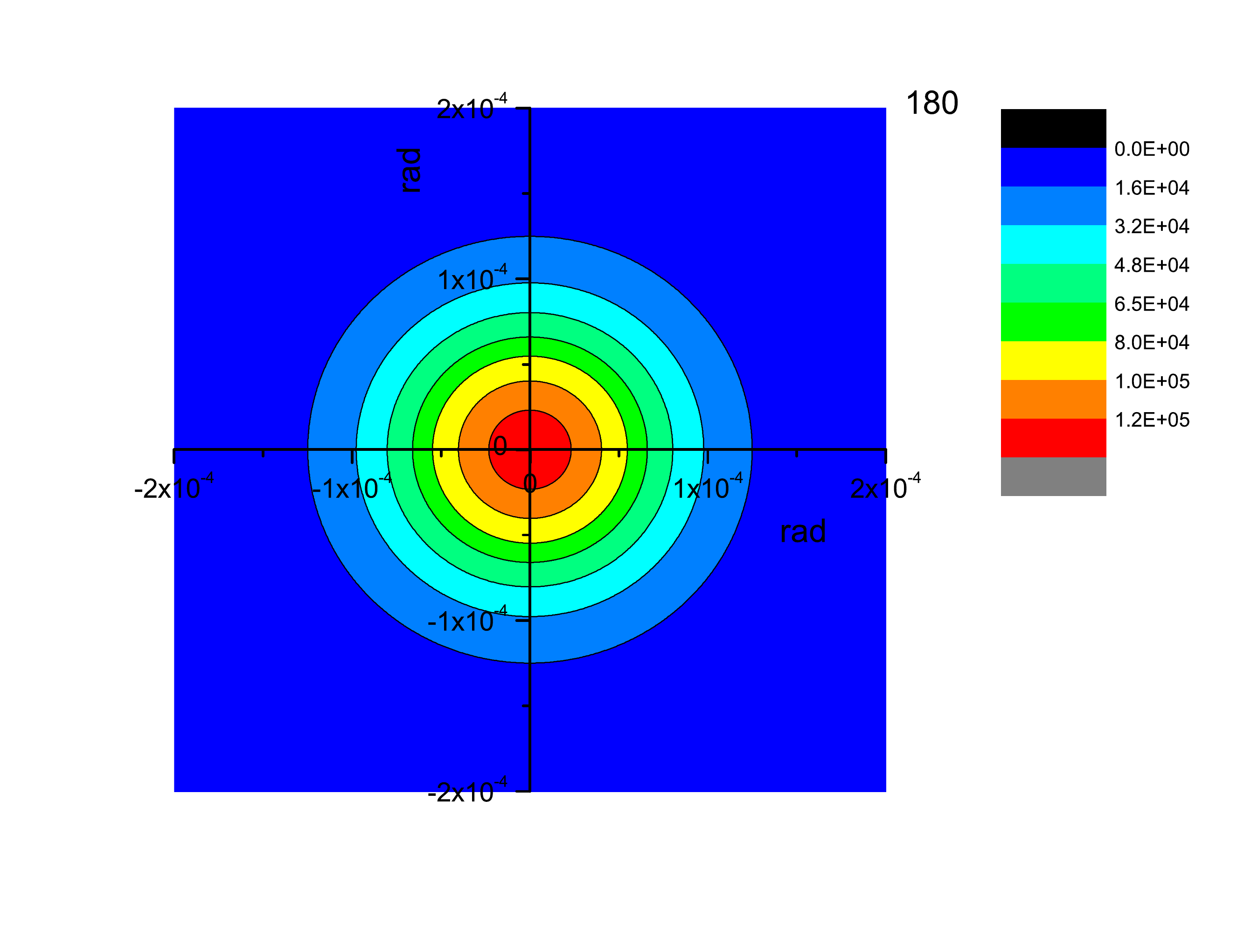} 
}
\caption{  
Pseudo-scalar generating differential cross sections through reaction $e^-+\gamma \rightarrow e^-+a$. 
The mass is chosen to be $0$. 
The origin represents the z-direction. 
 The exit angle $\theta$ of the EMPs represented by $\sqrt{x^2+y^2}$
The unit is $barn/(g^2_{ae}\cdot sr^2)$.
}
\label{fig.DCS.pScalar}
\end{figure*}

\subsection{dark photon}
\begin{figure*}
\centering
\subfigure{
\includegraphics[width=0.45\textwidth]{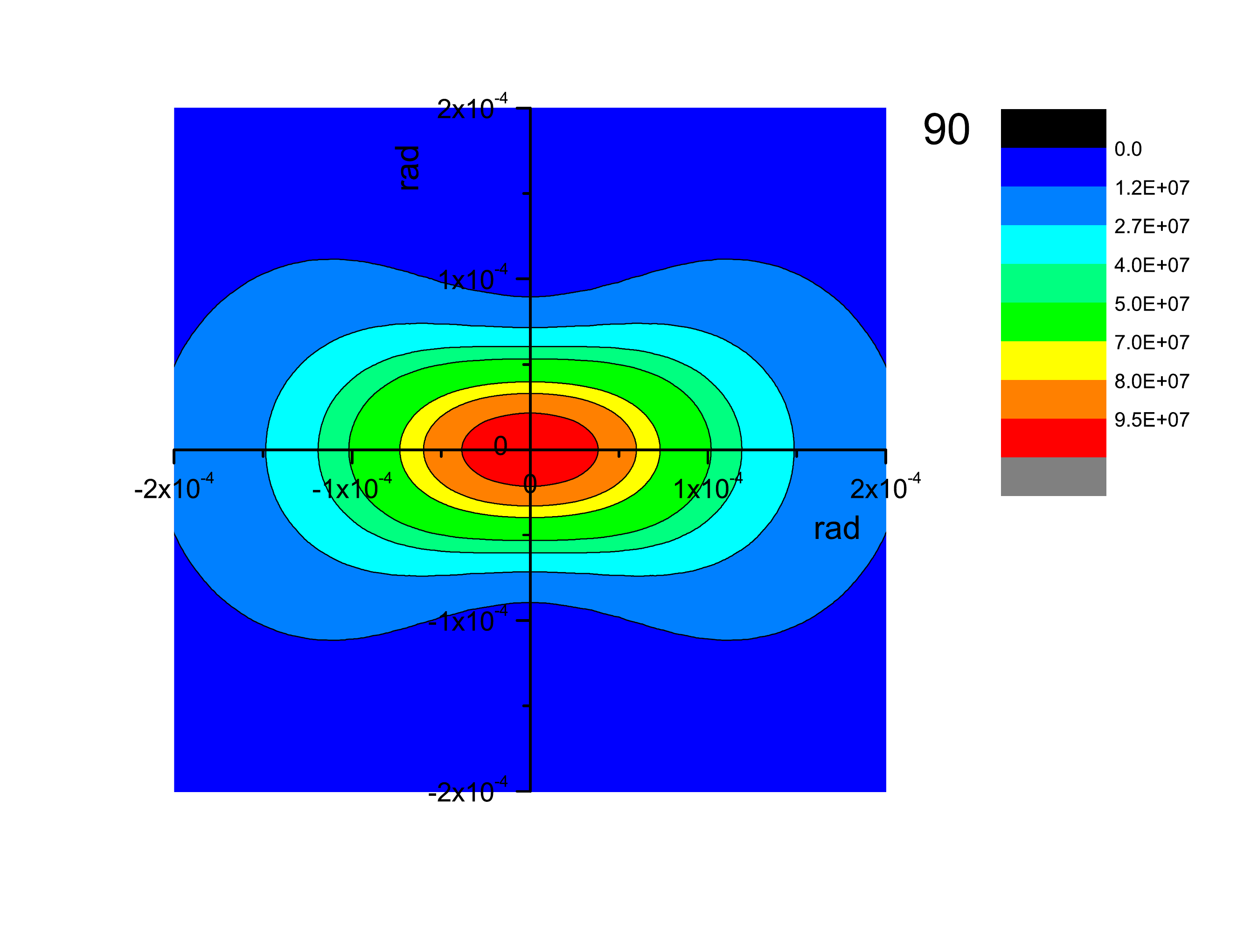} 
\includegraphics[width=0.45\textwidth]{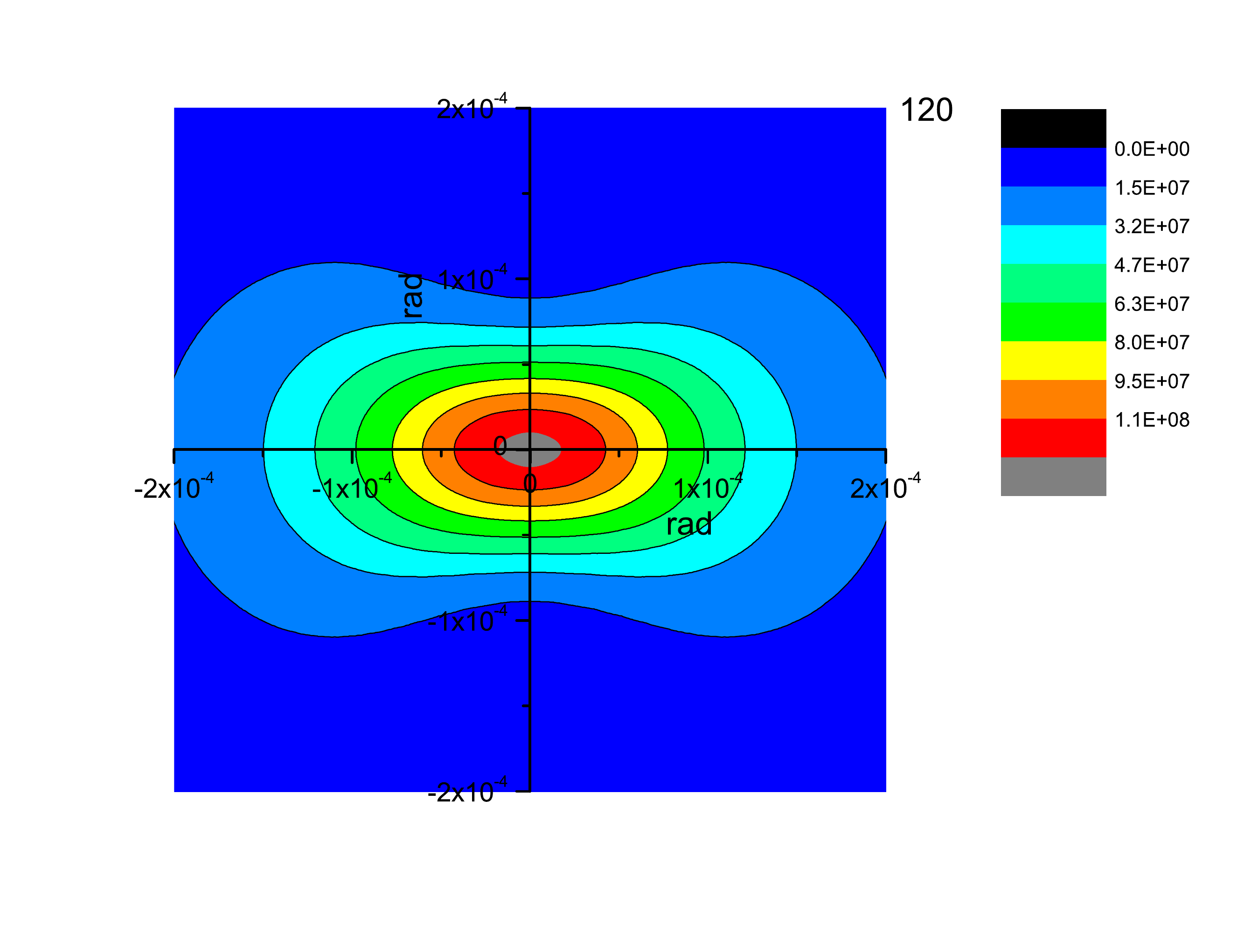} 
}
\subfigure{
\includegraphics[width=0.45\textwidth]{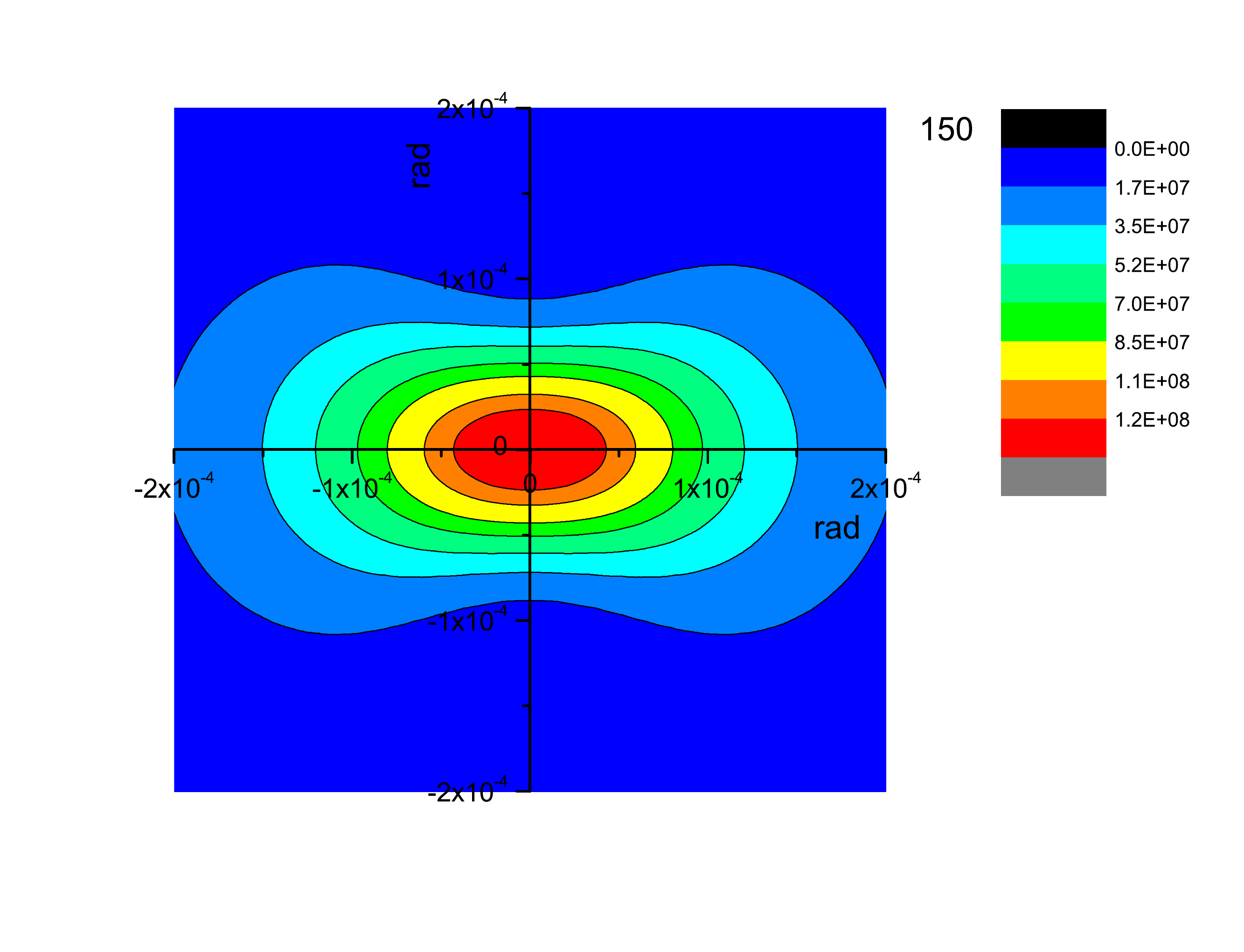} 
\includegraphics[width=0.45\textwidth]{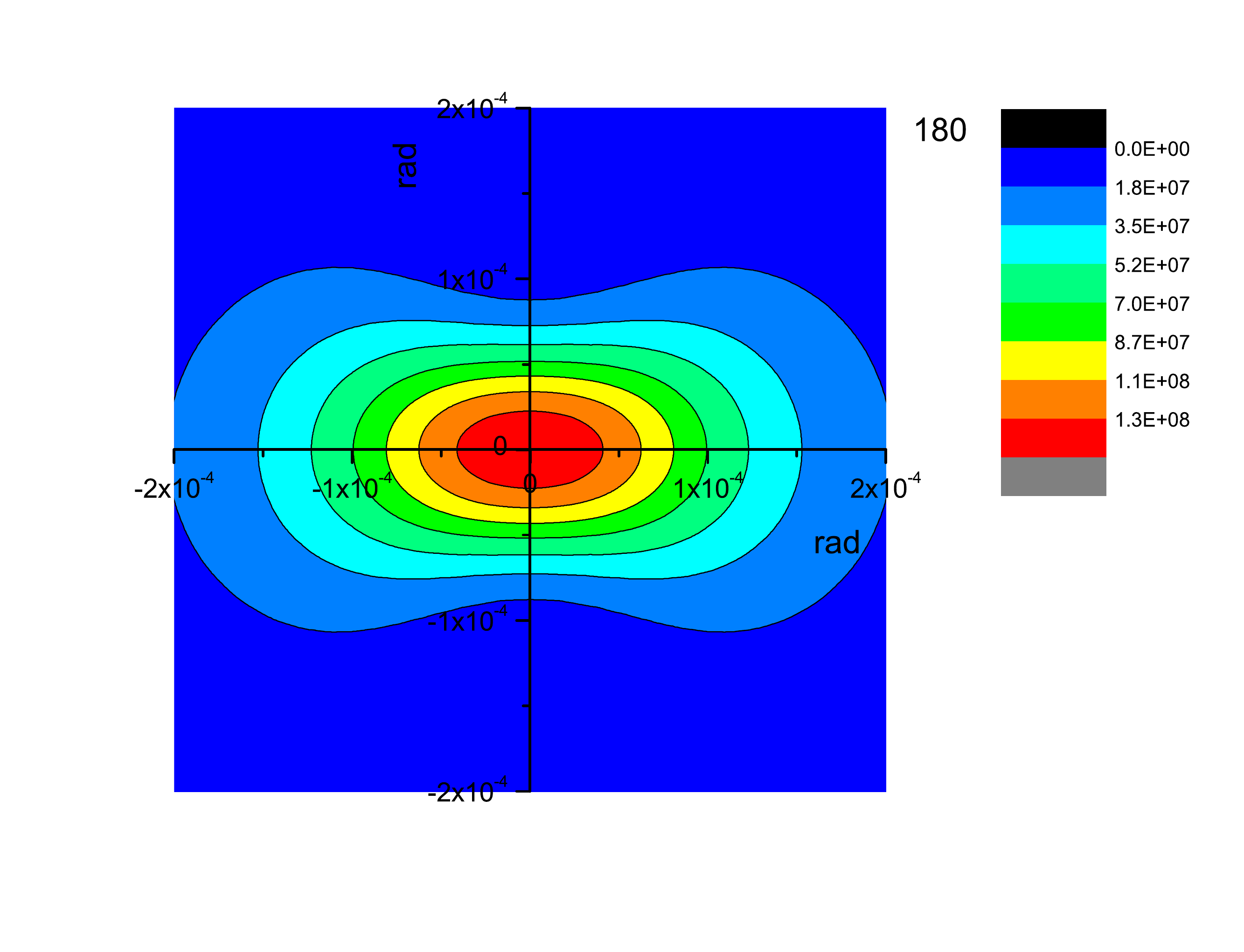} 
}

\caption{  
Dark photon generating cross sections through reaction $e+\gamma \rightarrow e+A'$. 
The mass is chosen to be $0$.  
The origin represents the z-direction. 
 The exit angle $\theta$ of the EMPs represented by $\sqrt{x^2+y^2}$
The unit is  $barn/(g^2_{A'e}\cdot sr^2)$.
}
\label{fig.DCS.DarkPhothon}
\end{figure*}
Similar to the psudo scalar case, the $\frac{1}{2}\sum_{spin}|M|^2$ of the vector (dark photon ) can be written as,
\begin{equation}
iM=\bar{u}(p') (-i g_{A'e}\epsilon^*_\nu(k') \gamma^\nu )\frac{i({p\mkern-7.5mu/}+{k\mkern-7.5mu/}+m_e)}{(p+k)^2-m_e^2} (-ie\epsilon_\mu(k)\gamma^\mu)u(p)+\bar{u}(p')(-ie\epsilon_\mu(k)\gamma^\mu) \frac{i({p\mkern-7.5mu/}-{k\mkern-7.5mu/}'+m_e)}{(p-k')^2-m_e^2} 
(-i g_{A' e} \epsilon^*_\nu(k') \gamma^\nu )u(p),
\end{equation}
where $\epsilon_\nu$ is the polarization direction of the photon. 
In this case, it points to the y-direction. The nominators are    
\begin{eqnarray}
I&=&8(p'\cdot k)(p\cdot k)-16( p\cdot k)(p' \cdot \epsilon)( p\cdot \epsilon)+16( p'\cdot k)( p\cdot \epsilon)( p\cdot \epsilon)+16( p'\cdot p)( p\cdot \epsilon)( p\cdot \epsilon)-32m_e^2 ( p\cdot \epsilon )( p\cdot \epsilon)\\
\frac{II}{2}&=&-8( p'\cdot \epsilon)( k\cdot p)( p\cdot \epsilon)+8( p'\cdot k)( p\cdot \epsilon)( p\cdot \epsilon)+8( p'\cdot k)( p\cdot \epsilon)( p'\cdot \epsilon)-8( p'\cdot \epsilon)( k\cdot p)(p' \cdot \epsilon)\\
&&+16( p'\cdot p)(p' \cdot \epsilon)( p\cdot \epsilon)-32m_e^2(p' \cdot \epsilon)( p\cdot \epsilon)\\
III&=&8(p' \cdot k)( p\cdot k)-16(k \cdot p)(p' \cdot \epsilon)( p'\cdot \epsilon)+16( p'\cdot k)(p \cdot \epsilon)( p'\cdot \epsilon)+16( p'\cdot p)( p'\cdot \epsilon)( p'\cdot \epsilon)\\
&&-32m_e^2 ( p'\cdot \epsilon)( p'\cdot \epsilon)
\end{eqnarray} 
The differential cross section at different collision angle 
$\theta=90^\circ, 120^\circ, 150^\circ$, and $180^\circ$ are shown in Fig.\ref{fig.DCS.DarkPhothon}. 
As expected, because the energy of the electron is much larger than that of the photon, 
in the lab frame, the expected scalar DMPs are highly concentrated in the forward angle.
This property will benefit experimental detection.
 
\subsection{scalar}
\begin{figure*}
\centering
\subfigure{
\includegraphics[width=0.45\textwidth]{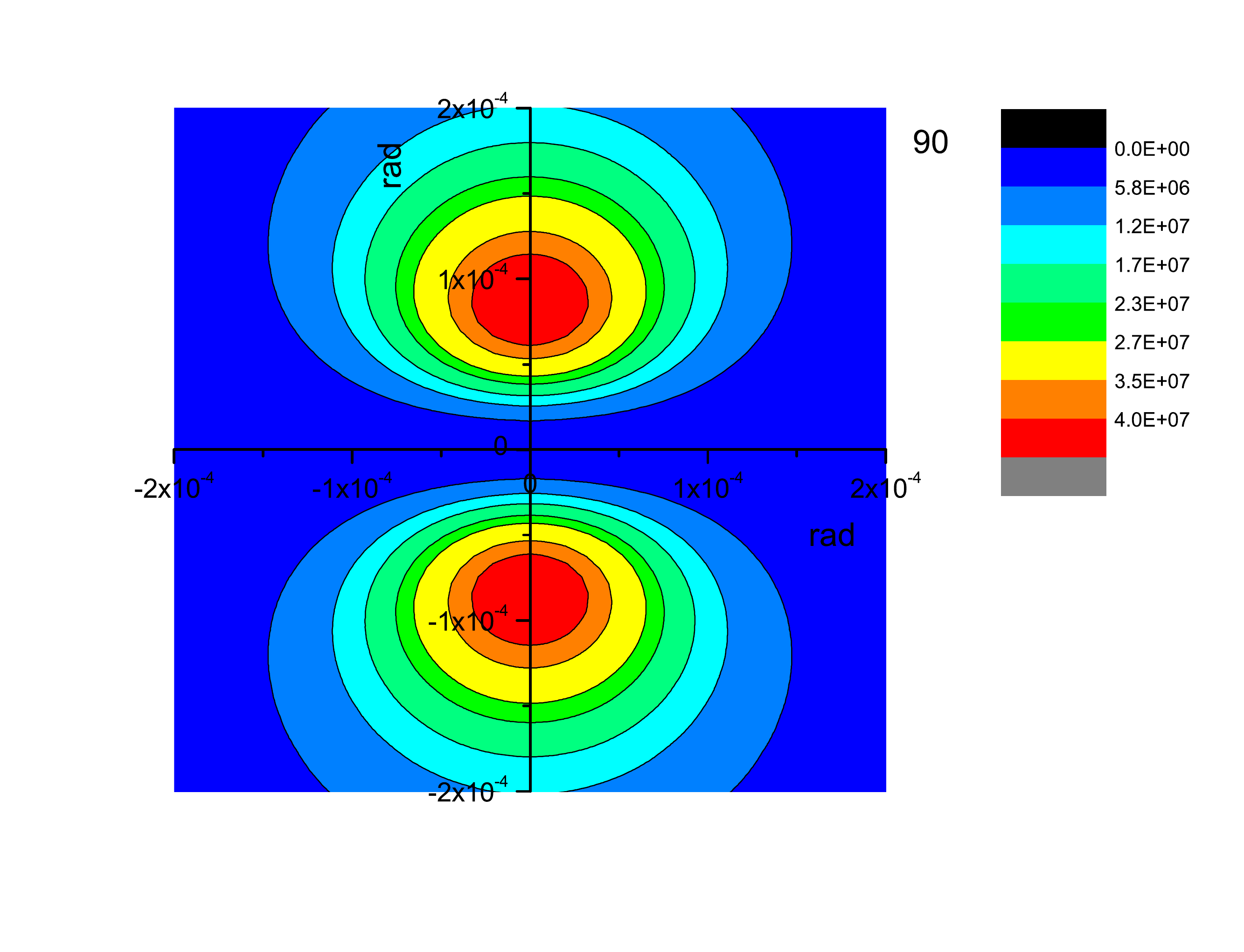} 
\includegraphics[width=0.45\textwidth]{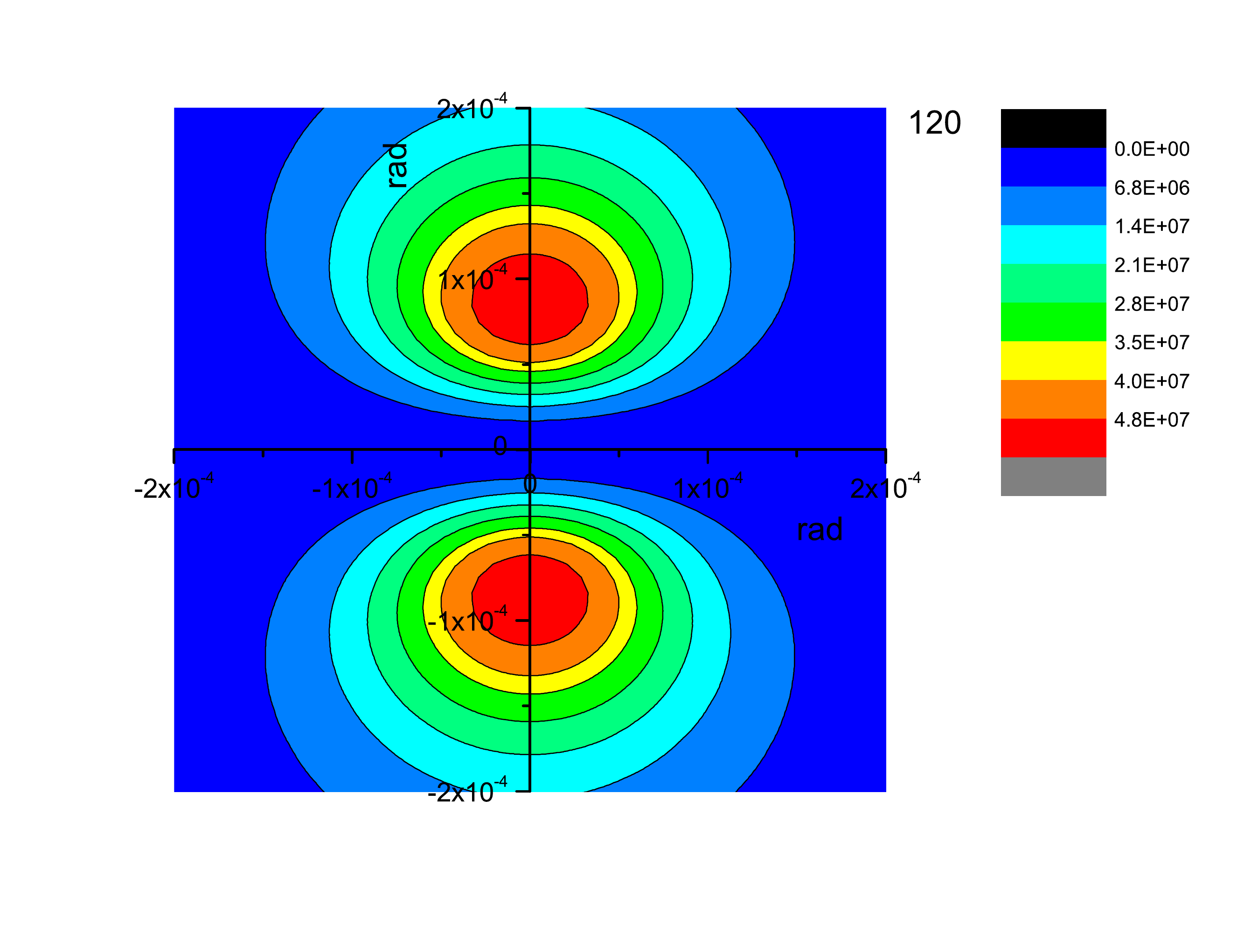} 
}
\subfigure{
\includegraphics[width=0.45\textwidth]{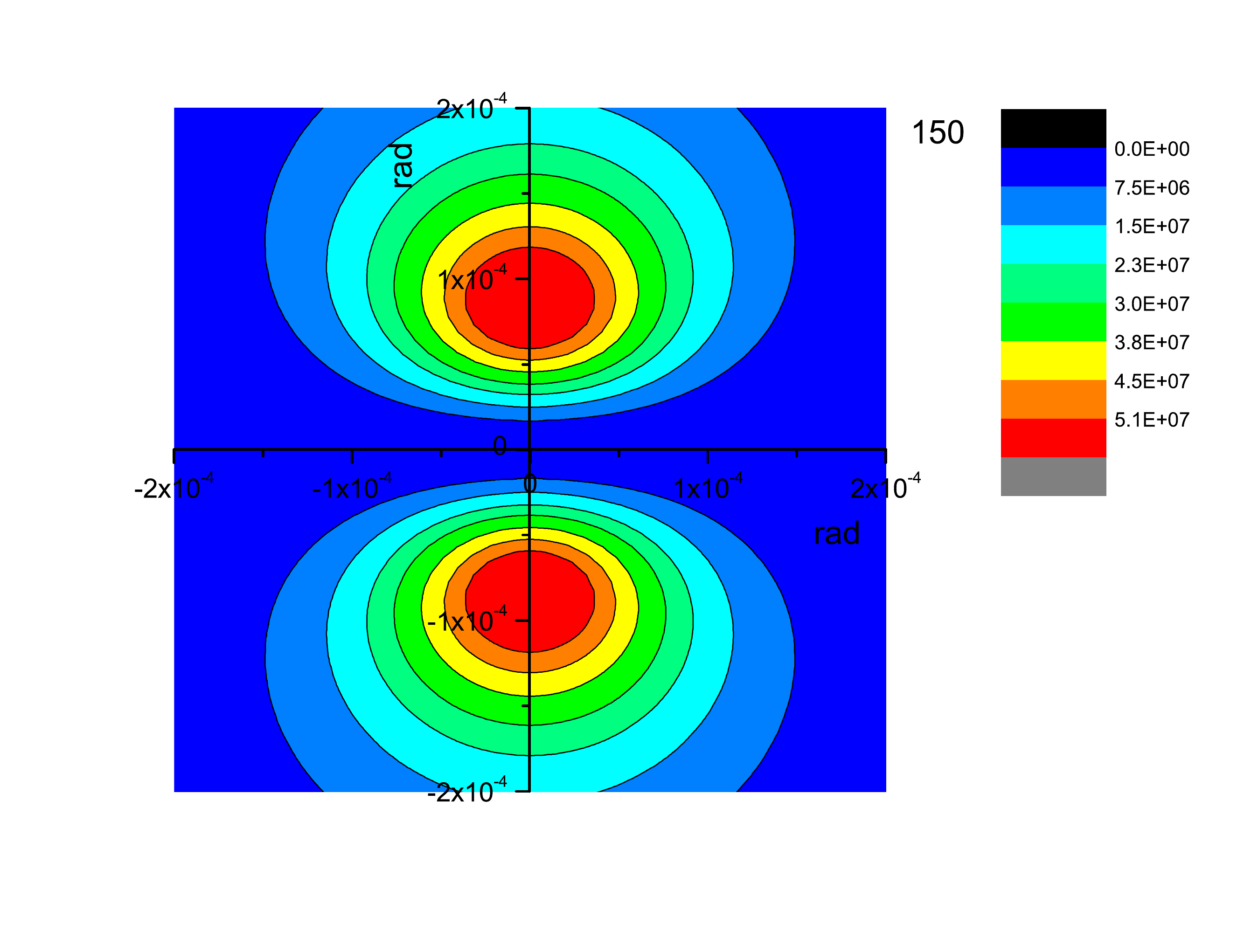} 
\includegraphics[width=0.45\textwidth]{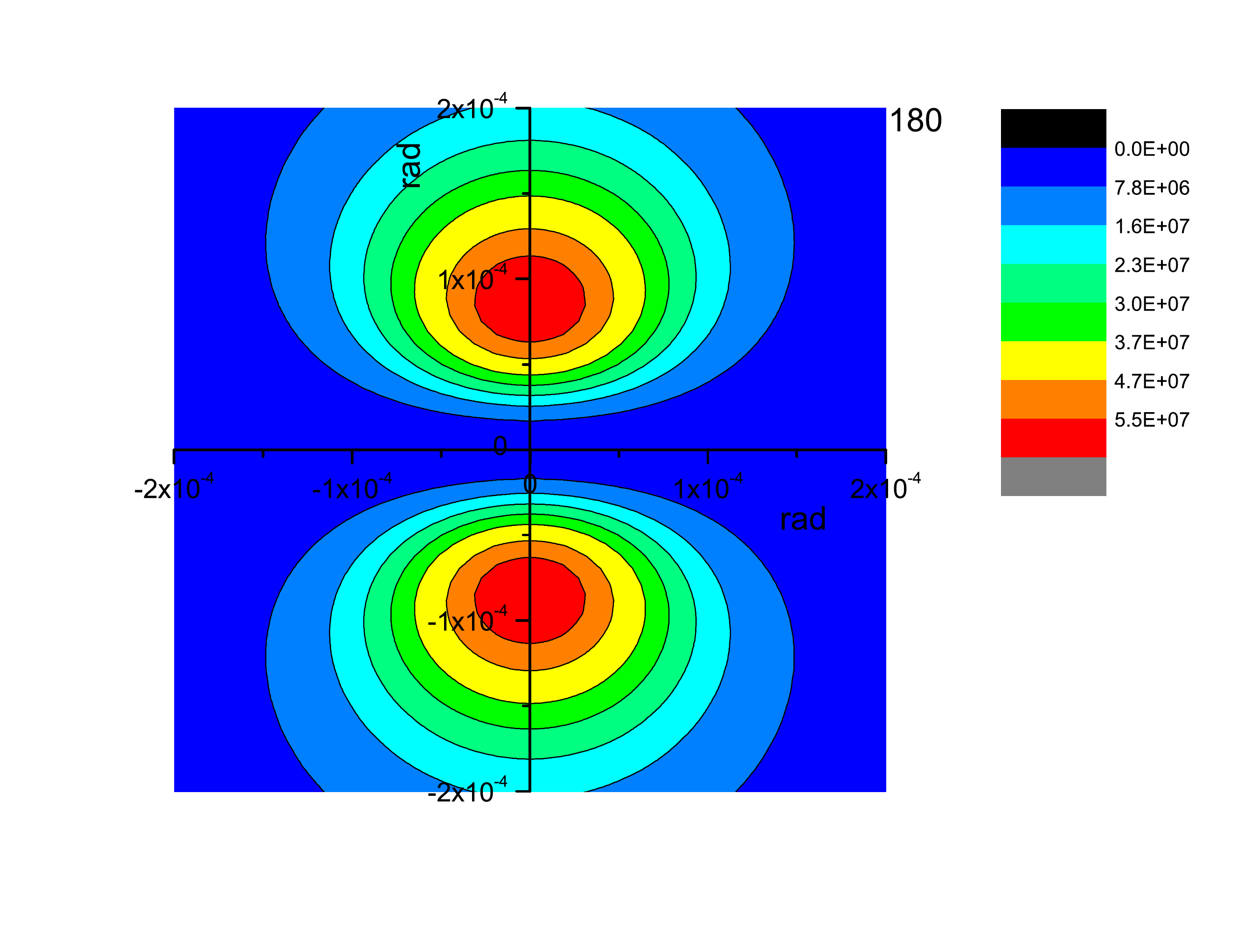} 
}

\caption{  
Scalar generating cross sections through reaction $e+\gamma \rightarrow e+\phi$. 
The mass is chosen to be $0$. 
The origin represents the z-direction. 
The exit angle $\theta$ of the EMPs represented by $\sqrt{x^2+y^2}$
The unit is $barn/(g^2_{\phi e}\cdot sr^2)$.
}
\label{fig.DCS.scalar}
\end{figure*}

Similar to the psudo-scalar case, the $\frac{1}{2}\sum_{spin}|M|^2$ of the scalar particles can be written as,
\begin{equation}
iM=\bar{u}(p') (-i g_{\phi e} )\frac{i({p\mkern-7.5mu/}+{k\mkern-7.5mu/}+m_e)}{(p+k)^2-m_e^2} (-ie\epsilon_\mu(k)\gamma^\mu)u(p)+\bar{u}(p')(-ie\epsilon_\mu(k)\gamma^\mu) \frac{i({p\mkern-7.5mu/}-{k\mkern-7.5mu/}'+m_e)}{(p-k')^2-m_e^2} (-i g_a )u(p),
\end{equation}
In this case it point to the y-direction. The nominators are  
\begin{eqnarray}
I&=&4( p'\cdot k)( p\cdot k)+8( p\cdot \epsilon)( p'\cdot k)(\epsilon \cdot p)-8( p\cdot \epsilon)( p'\cdot \epsilon)( p\cdot k)+8(p \cdot \epsilon)( p\cdot \epsilon)( p'\cdot p)+8m_e^2( p\cdot \epsilon)( p\cdot\epsilon )\\
\frac{II}{2}&=&4( p'\cdot k)( k\cdot p)-4(p \cdot \epsilon)( p'\cdot \epsilon)( k\cdot p)+4( p\cdot \epsilon)( p'\cdot k)(p \cdot \epsilon)-4( p'\cdot \epsilon)(p' \cdot \epsilon)(p \cdot k)+4(p' \cdot \epsilon)( p'\cdot k)(p \cdot \epsilon)\\
&&+8( p'\cdot p)(p' \cdot \epsilon)( p\cdot \epsilon)+8m_e^2(p' \cdot \epsilon)(p \cdot \epsilon)\\
III&=&4(p \cdot k)( p'\cdot k)-8( p'\cdot\epsilon )( p'\cdot \epsilon)( k\cdot p)+8(p' \cdot \epsilon)( p'\cdot k)( p\cdot \epsilon)+8( p'\cdot \epsilon)( p'\cdot \epsilon)( p'\cdot p)+8m_e^2 (p' \cdot \epsilon)(p' \cdot \epsilon)
\end{eqnarray}
where $\epsilon$ is the polarization direction of the photon. 
The differential cross section at different collision angle 
$\theta=90^\circ, 120^\circ, 150^\circ$, and $180^\circ$ are shown in Fig.\ref{fig.DCS.scalar}. 
As expected, because the energy of the electron is much larger than that of the photon, 
in the lab frame, the expected scalar particles are highly concentrated in the forward angle.
This property will benefit experimental detection.

%

\end{document}